\documentclass[aps,prb,twocolumn,showpacs,superscriptaddress,10pt,a4paper]{revtex4-1}
\usepackage{graphicx}
\usepackage{amsmath}
\usepackage{amssymb}
\usepackage{verbatim}
\usepackage{color}
\usepackage{dsfont}
\usepackage{ulem}

\newcommand{\m}[1]{\mathrm{#1}}

\usepackage{amsthm}

 \theoremstyle{definition}

 \theoremstyle{remark}

\begin{document}

\title{Majorana states in inhomogeneous spin ladders}

\author{Fabio L. Pedrocchi}
\affiliation{Department of Physics, University of Basel, Klingelbergstrasse 82, CH-4056 Basel, Switzerland}

\author{Stefano Chesi}
\affiliation{Department of Physics, McGill University, Montreal, Quebec, Canada H3A 2T8}

\author{Suhas Gangadharaiah}
\affiliation{Department of Physics, University of Basel, Klingelbergstrasse 82, CH-4056 Basel, Switzerland}
\affiliation{Indian Institute of Science Education and Research, Bhopal 462 023, India}

\author{Daniel Loss}
\affiliation{Department of Physics, University of Basel, Klingelbergstrasse 82, CH-4056 Basel, Switzerland}

\begin{abstract}

We propose an inhomogeneous open spin ladder, related to the Kitaev honeycomb model, which can be tuned between topological and nontopological phases. 
In extension of Lieb's theorem, we show numerically that the ground state of the spin ladder is either vortex free or vortex full.
We study the robustness of Majorana end states (MES) which emerge at the boundary between sections in different topological phases and show that while the MES in the homogeneous ladder are destroyed by single-body perturbations, in the presence of inhomogeneities at least two-body perturbations are required to destabilize MES. Furthermore, we prove that $x$, $y$, or $z$ inhomogeneous magnetic fields are not able to destroy the topological degeneracy. Finally, we present a trijunction setup where MES can be braided. A network of such spin ladders provides thus a promising platform for realization and manipulation of MES.
\color{black}
\end{abstract}


\maketitle

\section{Introduction}
The study of Majorana fermions in various solid-state systems has recently attracted a lot of attention. \cite{Kitaev2001, KitaevHoney, Bravyi2006, KaneGeneral2008, NayakGeneral2008,Terhal2012,AliceaWire2010, LutchynWire2010, OregWire2010, Tserkovnyak2011, ShankarWire2010, AliceaWire2011, Vishveshwara2011, KlinovajaWire2011, SuhasWire2011, MirceaWire2012,Sato1,Sato2,Stano2012} In particular, the possibility of realizing them as zero-energy states localized at the end of one-dimensional systems [so-called Majorana end states (MES)] has been the subject of many recent investigations. \cite{AliceaWire2010, LutchynWire2010, OregWire2010, AliceaWire2011, Tserkovnyak2011, ShankarWire2010, Vishveshwara2011, KlinovajaWire2011,SuhasWire2011,MirceaWire2012,Sato1,Sato2,Stano2012} 
Besides being of fundamental interest, the study of MES is motivated by their potential use for topological quantum computing. 

While electronic systems are a natural choice for the realization of such MES, well known fermionization techniques have stimulated the study of MES appearing in the Fermionic mapping of one-dimensional spin systems. \cite{Tserkovnyak2011,Vishveshwara2011,ShankarWire2010} Since many proposals exist for implementing designer spin-spin interactions, \cite{Weimer2010} and, in particular, for the experimental realization of Kitaev-like spin models, \cite{realization1,realization2,realization3} such one-dimensional systems are interesting candidates \color{black} for the realization and detection of MES. 

Proposals of highly engineered spin interactions also suggest the implementations of inhomogenous systems where regions in different topological phases coexist, which is the main feature of the spin ladder studied in this work. The spin ladder is based on nearest-neighbor Ising interactions, in extension of the Kitaev honeycomb model,\cite{KitaevHoney} and consists of sections which alternate between the topological and nontopological phases. The phase of each section is simply determined by the strength of the Ising couplings and at the  boundary between different phases (as well as at the open ends of the ladder) single well-localized MES exist. Being interested in the ground-space properties of the model, we perform an extensive numerical analysis of possible vortex configurations. We always find that the ground state is either vortex-free or vortex-full and thereby give strong evidence that that the Lieb theorem, \cite{Lieb1994} originally formulated for different boundary conditions, also applies to such open spin ladders. While freely moving vortices would introduce additional degeneracies, the presence of a finite gap for the vortex excitations makes it possible to characterize the ground space in terms of spatially localized Majorana states, and to consider braiding of such MES. We discuss a trijunction setup similar to the one proposed in Ref.~\onlinecite{AliceaWire2011}, which could allow the implementation of topologically protected  quantum gates.

In contrast to previous studies,\cite{Tserkovnyak2011,Vishveshwara2011,ShankarWire2010} we do not analyze our model by making use of the Jordan-Wigner transformation, but consider the alternative mapping proposed by Kitaev.\cite{KitaevHoney} This method has the interesting feature of mapping local spin operators to local Fermionic operators and has some advantage in analyzing the stability of the MES. In particular, it makes immediately clear that MES in the homogeneous ladder are already fragile against single-body perturbations. This conclusion is consistent with general arguments showing that topological protection in one-dimensional systems can only be effective against local perturbations conserving certain symmetries.\cite{Verstraete2005,Wen2011,Schuch2011} In $p$-wave wires, for example, MES are protected against local perturbations which conserve parity. Such symmetry assumption is realistic for superconducting wires (although other limitations exist due to quasiparticle poisoning \cite{Rainis2012}) but \color{black} an analogous constraint does not appear for general spin systems. For this reason we study in great detail the robustness of MES and characterize the form of local perturbations which can split the corresponding degeneracy. In the Fermionic language adopted here, the fragility of the topological degeneracy results from the only apparent locality of the Kitaev mapping. Non-local strings of Fermionic operators coupling the MES appear through the projector onto the physical subspace.\cite{KitaevHoney,Chesi2011} Nevertheless, we show that the inhomogenous ladder has better properties than the homogenous ladder \cite{Vishveshwara2011} since single-body perturbations are not sufficient to split the degeneracy but at least two-body interactions are necessary. Furthermore, we show that the topological degeneracy of our model is robust against inhomogeneous magnetic fields along $x$, $y$, or $z$.

The paper is organized as follows. In Sec. \ref{sec:1} we introduce the inhomogeneous ladder and the spin-to-fermion mapping we use to study MES properties. In Sec. \ref{sec:2}, we derive the topological invariant of the spin ladder for both the vortex-free and the vortex-full sectors. We show that well-localized MES appear at the junction between topological and nontopological sections and demonstrate how to move MES along the ladder. In Sec. \ref{sec:3} we investigate numerically the validity of the Lieb theorem \cite{Lieb1994} and show that it also applies to this type of open spin ladders, since the ground state is either vortex-free (when one of the coupling is negative) or vortex-full (when all couplings are positive). In Sec. \ref{sec:4}, we perform an analysis of the robustness of MES under local perturbations. We demonstrate that single-body perturbations split the topological degeneracy of the homogeneous ladder, while two-body terms are required in the inhomogeneous case. Furthermore, we prove that $x$, $y$, or $z$ inhomogeneous magnetic fields alone are not able to split the topological degeneracy of the ladder. In Sec. \ref{sec:5} we present a trijunction setup which could allow braiding of MES following the scheme of Ref.~\onlinecite{AliceaWire2011} and Sec. \ref{sec:6} contains our final remarks. We present additional details and interesting aspects of the model in Appendixes \ref{sec:twowires}-\ref{sec:correlations}.

\section{Inhomogeneous spin ladder and Kitaev's mapping}\label{sec:1}
\begin{figure}[h]
	\centering
		\includegraphics[width=0.45\textwidth]{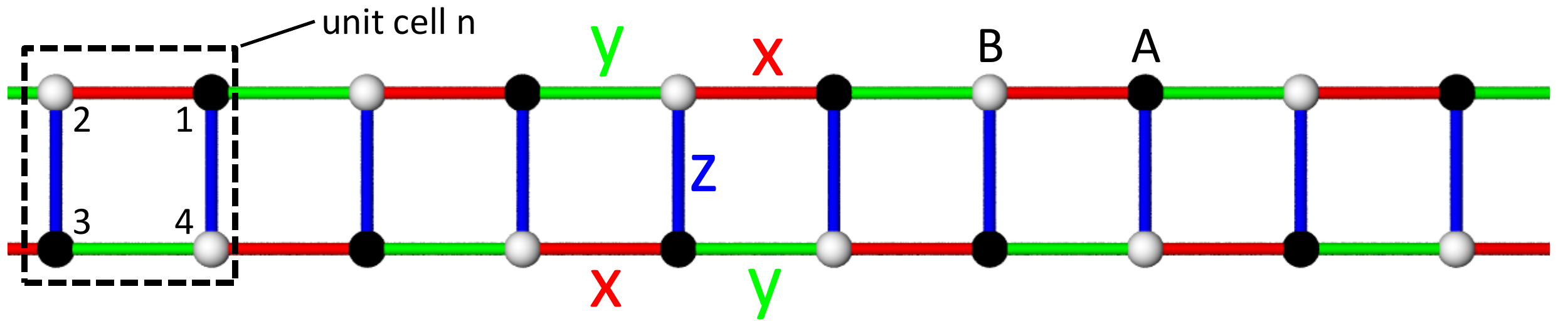}
	\caption{Inhomogeneous spin ladder. Each site contains a quantum spin-$\frac{1}{2}$ which interacts with its nearest-neighbor spins via bond-dependent Ising interactions $J_{x}, J_{y}, J_{z}$.
	The $x$, $y$, and $z$ bonds are indicated by red, green, and blue lines, respectively, the $A$ ($B$) sublattice site by black (white) dots, and the n{\it th} unit cell composed of four spins by the black dashed square. In contrast to the standard honeycomb model, the $z$ couplings are inhomogeneous, {\it i.e.}, site-dependent $J_{z}\rightarrow J_{z_{ij}}$.}
	\label{fig:Chain_2}
\end{figure}
The spin model we propose is an inhomogeneous ladder version of the compass \cite{KugelKhomskii82} or Kitaev honeycomb \cite{KitaevHoney} model with Hamiltonian
\begin{equation}\label{eq:Hamiltonian}
H=\sum_{\langle i,j\rangle}J_{\alpha_{ij}}\sigma_{i}^{\alpha_{ij}}\sigma_{j}^{\alpha_{ij}},
\end{equation}
where $\boldsymbol{\sigma}_{i}=({\sigma}_{i}^x,{\sigma}_{i}^y,{\sigma}_{i}^z)$ are the Pauli operators for the spin-1/2 located at the site $i$ of the ladder, and where the sum runs over all pairs of nearest-neighbor sites of the open ladder containing $N$ unit cells.
We assume that the ladder is of length $2N-1$ (lattice constant set to one) , {\it i.e.} consists of an odd number $2N-1$ of square plaquettes, see Fig.~\ref{fig:Chain_2}.
The anisotropy direction in spin space of the Ising interaction $J_{\alpha_{ij}}$ depends on the orbital location of the bond which is labeled by the index $\alpha_{ij}=x,y,z$ for $x$, $y$, and $z$-bonds, respectively, see Fig.~\ref{fig:Chain_2}.
Furthermore, we allow the $z$ couplings to depend on position, {\it i.e.}, $J_{z}\rightarrow J_{z_{ij}}$. 
Without loss of generality we assume that $J_{z_{ij}}>0$. 
Following Ref.~\onlinecite{KitaevHoney}, 
this model can be solved exactly in an extended Hilbert space $\widetilde{\mathcal{L}}$ by assigning four Majorana fermion operators $b_{i}^{x,y,z}$ and $c_{i}$ (all self-adjoint) to each site of the lattice and mapping each spin operator to a product of two Majoranas, 
\begin{equation}\label{eq:mapping}
\widetilde{\sigma}_{i}^{\alpha}=ib_{i}^{\alpha}c_{i}.
\end{equation}
In fermionic representation, the spin Hamiltonian in Eq.~(\ref{eq:Hamiltonian}) takes the form $\widetilde{H}=i\sum_{\langle i,j\rangle}\widehat{A}_{ij}c_{i}c_{j}$,
where $\widehat{A}_{ij}=J_{\alpha_{ij}}\hat{u}_{ij}$ and $\hat{u}_{ij}=ib_{i}^{\alpha_{ij}}b_{j}^{\alpha_{ij}}$.
The $u$ operators commute with each other and with $\tilde{H}$, and satisfy $\hat{u}^2_{ij}=1$. Therefore, the extended Hilbert space splits into subspaces $\widetilde{\mathcal{L}}_{u}$, {\it i.e.}, $\widetilde{\mathcal{L}}=\oplus \widetilde{\mathcal{L}}_{u}$, where $u$ represents a certain configuration of eigenvalues $u_{ij}=\pm1$. 
To remove the ambiguity due to $\hat{u}_{ij}=-\hat{u}_{ji}$, we assume that for a chosen value ${u}_{ij}$ the first index $i$ belongs to sublattice A (see Fig.~\ref{fig:Chain_2}).
The physical subspace is defined through the gauge operators $D_{i}=b_{i}^{x}b_{i}^{y}b_{i}^{z}c_{i}$ as $\mathcal{L}=\{\vert\Psi\rangle\,\,:\,\,D_{i}\vert\Psi\rangle=\vert\Psi\rangle\}$. Starting from a state $\vert\Psi\rangle$ in the extended space, the corresponding physical state $\vert\Psi\rangle_{\m{phys}}$ is given by symmetrization over all gauge operators $D_{i}$, i.e.,\cite{KitaevHoney}
\begin{equation}\label{eq:symmetrization}
\vert\Psi\rangle_{\m{phys}}=\prod_{i=1}^{4N}\left(\frac{1+D_{i}}{2}\right)\vert\Psi\rangle.
\end{equation}
\color{black}
In each subspace $\widetilde{\mathcal{L}}^{u}$, the operators $\widehat{A}_{ij}$ are replaced by numbers $A_{ij}^{u}$ and thus the quadratic Hamiltonian $\widetilde{H}$ is easily solvable with a canonical transformation $Q^{u}$ to new Majorana modes \cite{KitaevHoney}
\begin{equation}\label{eq:canonical}
(b_{1},b_{2},...,b_{4N-1},b_{4N})=(c_{1},...,c_{4N})Q^{u}.
\end{equation}
Under this transformation, 
$\widetilde{H}$ takes, for a given configuration $u$, the form 
$\widetilde{H}_{u}=\frac{i}{2}\sum_{m=1}^{N}\epsilon_{m}b_{2m-1}b_{2m}$, where $\epsilon_{m}>0$ are the positive eigenvalues of $2iA^{u}$. By defining new complex fermion operators $a_{m}=(b_{2m-1}+ib_{2m})/2$, we finally obtain $\widetilde{H}_{u}=\sum_{m}\epsilon_{m}(a_{m}^{\dagger}a_{m}-1/2)$.

The spin ladder Eq.~(\ref{eq:Hamiltonian}) possesses as conserved quantities two types of plaquettes that are naturally associated with each unit cell $n$, {\it i.e.}, $W_{n}=-\sigma_{n,1}^{y}\sigma_{n,2}^{y}\sigma_{n,3}^{x}\sigma_{n,4}^{x}$ and $\overline{W}_{n}=-\sigma_{n,1}^{x}\sigma_{n+1,2}^{x}\sigma_{n+1,3}^{y}\sigma_{n,4}^{y}$, where each spin $\sigma_{n,\alpha}$ is labeled by the index $n$ for the unit cell and $\alpha=1,...,4$ for one of the four spins inside the unit cell (see Fig.~\ref{fig:Chain_2}). We say that a square plaquette $p$ carries a vortex if $W_{n}=-1$ for $p=2n-1$ and if $\overline{W}_{n}=-1$ for $p=2n$. In fermionic representation, the plaquettes are  products of $u$ operators around each unit cell. In the following, we study the robustness of MES against local perturbations and show that not all the degeneracies due to MES can be split by single-body perturbations. Additionally, we show that the topological degeneracy cannot be fully split by inhomogeneous $x$, $y$, or $z$ magnetic fields. We note that in the special case with $J_{z_{ij}}=J_{z}$ our model is equivalent to the usual honeycomb model studied in Refs.~\onlinecite{Vishveshwara2011} and \onlinecite{Feng2007}. As we show below, the Majorana degeneracy present in the homogeneous ladder is less protected in the sense that it is fully split by single-body local perturbations. Similar homogeneous Kitaev ladders have been studied in Refs.~\onlinecite{Feng2007} and \onlinecite{Motrunich2011}.\color{black}
\section{Topological phases of spin ladders}\label{sec:2}
\begin{figure}[h]
	\centering
		\includegraphics[width=0.45\textwidth]{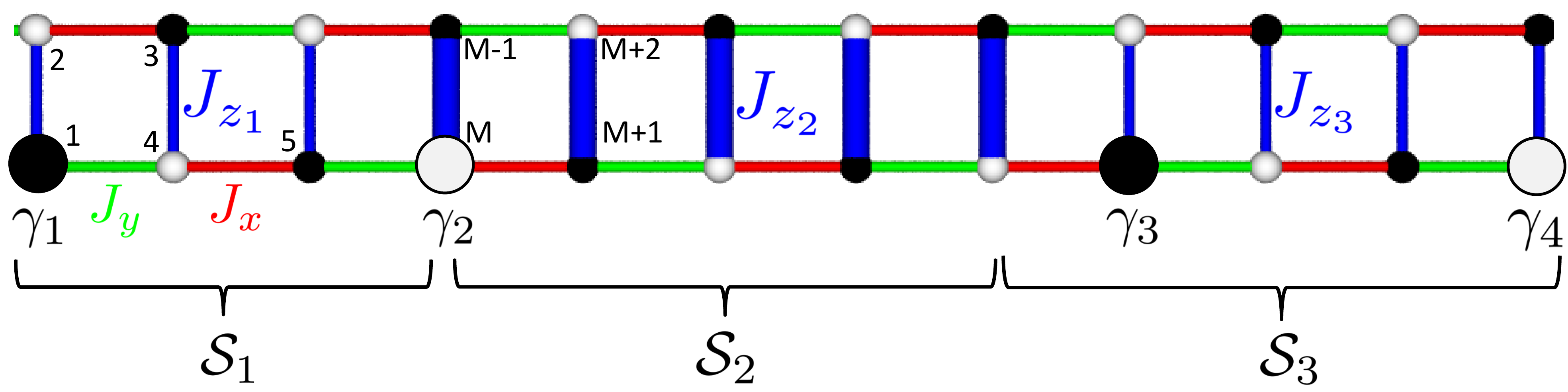}
	\caption{Inhomogeneous spin ladder with different topological sections. Shown are two topological sections $\mathcal{S}_{1}$ and $\mathcal{S}_{3}$ (thin $z$-bonds $J_{z_{1}}$ and $J_{z_{3}}$) separated by a nontopological section $\mathcal{S}_{2}$ (thick $z$-bonds $J_{z_{2}}$). For the corresponding
	$J_{x,y,z}$ values, see main text.
	The wave functions of the four MES $\gamma_{1,...,4}$ are mainly localized at the phase boundaries and, for $J_{x}>J_{y}$, on the lower sites of the ladder as indicated by the large dots.}
	\label{fig:Chain_5}
\end{figure}
We consider now a spin ladder with sections in different topological phases, $\mathcal{S}_{1}$, $\mathcal{S}_{2}$, and $\mathcal{S}_{3}$, which are distinguished by the value of the $J_{z_{ij}}$ couplings (see Fig.~\ref{fig:Chain_5}). If we focus on the vortex-free or vortex-full sector, then we choose the $J_{z_{ij}}$ couplings as follows: for the $\mathcal{S}_{1}$ and $\mathcal{S}_{3}$ parts, $J_{z_{ij}}=J_{z_{1}}=J_{z_{3}}$ and $\vert J_{z_{1}}\vert< \vert J_{x}\pm J_{y}\vert $, while for the $\mathcal{S}_{2}$ part we have $J_{z_{ij}}=J_{z_{2}}$ and $\vert J_{z_{2}}\vert> \vert J_{x}\pm J_{y}\vert $ (see Fig.~\ref{fig:Chain_5}). Below we derive the conditions for the existence of zero-energy MES in the vortex-free (vortex-full) sector with the use of the mapping (\ref{eq:mapping}) and prove that sections $\mathcal{S}_{1}$ and $\mathcal{S}_{3}$ are topological, while section $\mathcal{S}_{2}$ is nontopological. 

The vortex-free sector corresponds to all $u_{ij}=+1$. In contrast, the configuration where all the $u$'s along only one of the axes (say $x$ axis) take on the value $-1$ is vortex-full. From the explicit expression of $A_{ij}^{u}$ we obtain the bulk spectrum in the vortex-free sector for $J_{z_{ij}}=J_{z}$,
\begin{equation}\label{eq:spectrum1}
\epsilon_{{\stackrel{\scriptscriptstyle{{1+m}}}{ \scriptscriptstyle{{2+m}}}}}(k)=\pm2\sqrt{J^2+2J_{x}J_{y}\cos(k)-(1-m)
\gamma_{k}},
\end{equation}
where $J^2=J_{x}^2+J_{y}^2+J_{z}^2$, $\gamma_{k}=\sqrt{(2+2\cos(k))(J_{x}+J_{y})^2J_{z}^2}$, $k$ is the wave vector, and $m=0,2$.

In the vortex-free sector, the MES eigenvectors $\boldsymbol{\phi}$ with eigenvalues $\epsilon=0$ can be shown to satisfy the following transfer equations $(\phi_{n+1,\alpha+\xi},\phi_{n,\alpha+\xi})^{T}=\mathcal{T}_{\alpha}(\phi_{n,\alpha+\xi},\phi_{n-1,\alpha+\xi})^{T}$ (the two labels of $\phi_{n,\alpha}$ correspond to the unit cell $n$ and one site $\alpha$ of the unit cell),
with $\xi=0,2$, and $\mathcal{T}_{1,2}=\begin{pmatrix}\frac{J_{z}^2}{J_{x,y}^2}-\frac{2J_{y,x}}{J_{x,y}} & -\frac{J_{y,x}^2}{J_{x,y}^2}\\ 1 & 0\end{pmatrix}$. MES can exist only when both eigenvalues of $\mathcal{T}_{\alpha}$ have absolute value larger or smaller than one. Therefore, we define the topological invariants $\nu_{\alpha}=-\m{sgn}((1-\vert\tau_{1}^{\alpha}\vert)(1-\vert\tau_{2}^{\alpha}\vert))$,
where $\tau_{1,2}^{\alpha}$ are the two eigenvalues of $\mathcal{T}_{\alpha}$. The system is in the {\it nontopological phase} when $\nu=+1$, and in the {\it topological phase} with MES when $\nu=-1$. From the above explicit expression for $\mathcal{T}_{1,2}$, we obtain the following result for the topological invariants in agreement with Refs.~\onlinecite{Vishveshwara2011} and \onlinecite{Feng2007}
\begin{equation}\label{eq:nu}
\nu_{1}=\nu_{2}=\m{sgn}(2\vert J_{z}\vert-2\vert J_{x}+J_{y}\vert).
\end{equation}
In the vortex-full sector, the topological invariant is given by Eq.~(\ref{eq:nu}) with $J_{x}+J_{y}$ replaced with $J_{x}-J_{y}$. From Eq.~(\ref{eq:nu}) it follows that sections $\mathcal{S}_{1}$ and $\mathcal{S}_{3}$ of our model are topological, while section $\mathcal{S}_{2}$ lies in the nontopological phase. This system thus contains four $c$ Majoranas: $\gamma_{1,4}$ localized at each end of the ladder and $\gamma_{2,3}$ at each junction between topological and nontopological sections of the ladder, see Fig.~\ref{fig:Chain_5}. Note that for $J_{x}>J_{y}$ ($J_{y}<J_{x}$), $\gamma_{1}$ and $\gamma_{3}$ will reside on the $A$ ($B$) sublattice, while $\gamma_{2}$ and $\gamma_{4}$ on the $B$ ($A$) sublattice. For the rest of this work we always consider the case where $\vert J_{x}\vert>\vert J_{y}\vert$ and $J_z>0$, since the case $\vert J_{x}\vert<\vert J_{y}\vert$ can be treated analogously. \color{black}
It is worth pointing out that $\mathcal{T}_{1}=\mathcal{T}_{2}$ when $\vert J_{x}\vert=\vert J_{y}\vert$, and consequently all the $\vert\phi_{n,\alpha=1,...,4}\vert$ will decrease (increase) with $n$ if both eigenvalues of $\mathcal{T}_{1,2}$ have their absolute values smaller (larger) than one. This excludes the presence of MES localized at the right (left) end of the ladder. Consequently, Eq. (\ref{eq:nu}) is strictly valid only for $\vert J_{x}\vert\neq \vert J_{y}\vert$.  

\begin{figure}[h]
	\centering
		\includegraphics[width=0.44\textwidth]{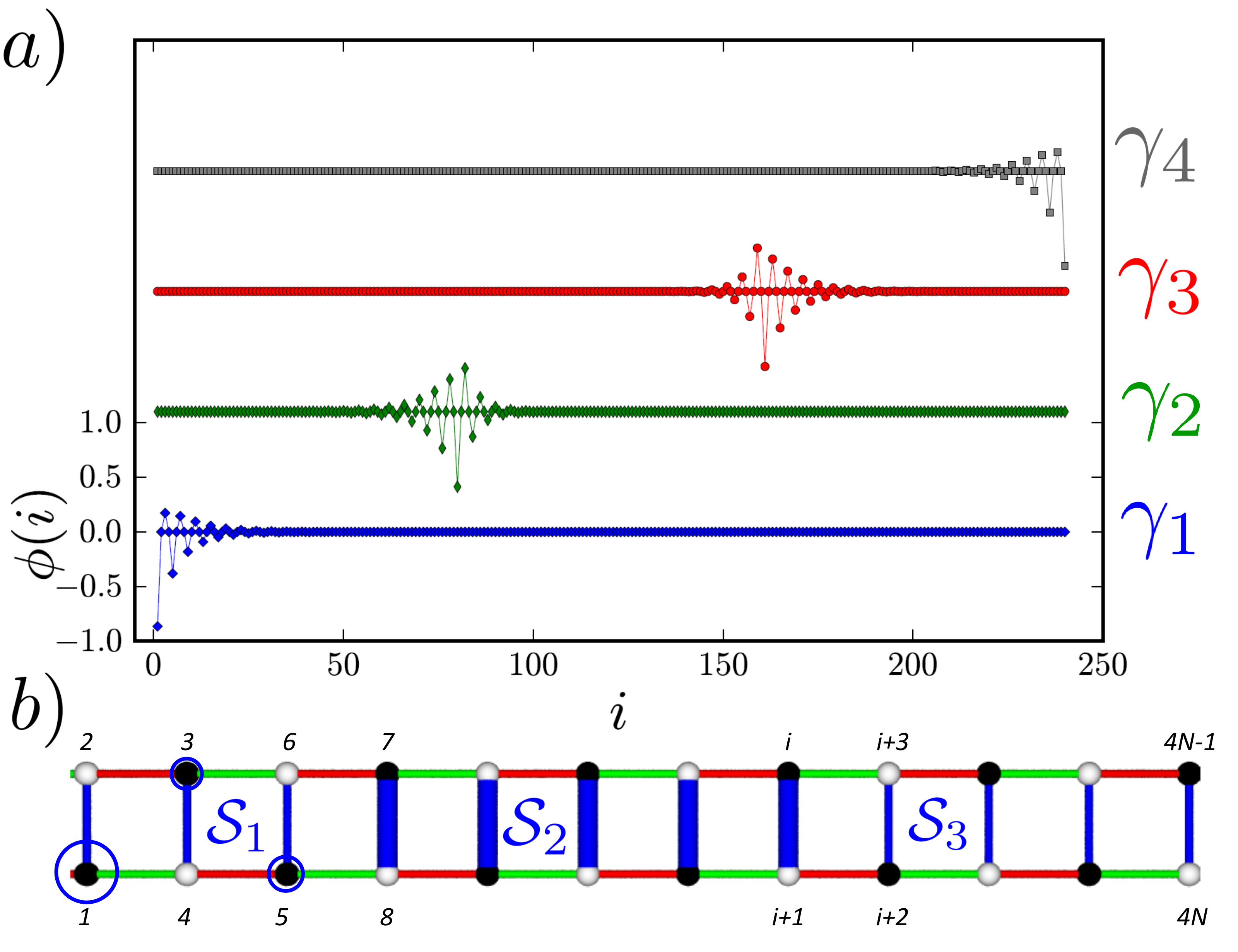}
	\caption{Inhomogeneous spin ladder as defined in Fig.~(\ref{fig:Chain_5}). $a)$ MES wave functions $\phi(i)$ (corresponding to $\gamma_{1,...,4}$) as functions of site $i$. The curves for $\gamma_{2,3,4}$ are shifted vertically for clarity. The order used for the site labeling of the spin ladder is shown in $b)$. The circles represent the wave function weight of $\gamma_{1}$ (proportional to the area enclosed by the circle) at the corresponding site. For both plots we have all $u_{ij}=+1$ (vortex-free), $N=60$, $J_{x}=1.0$, $J_{y}=-0.4$, $J_{z_{1}}=J_{z_{3}}=0.2$ in $\mathcal{S}_{1,3}$, and $J_{z_{2}}=3$ in $\mathcal{S}_{2}$. Section $\mathcal{S}_{2}$ starts at unit cell $n=41$ and ends at $n=79$.}
	\label{fig:Majorana_S_W}
\end{figure}

The four MES operators $\gamma_{1,...,4}$ can be easily expressed in terms of Majorana operators $c_{i}$ through relation (\ref{eq:canonical}): $\gamma_{1,...,4}=\sum_{j}Q_{i_{1,...,4},j}^{u}c_{j}$,
where the coefficients $Q_{i_{1 (2)},j}^{u}$ and $Q_{i_{3 (4)},j}$ are the elements of the imaginary (real) part of the $\epsilon=0$ eigenvectors of matrix $iA^{u}$. Figure~\ref{fig:Majorana_S_W} shows a plot of the wave functions of four Majoranas $\gamma_{1,...,4}$ for a ladder with $N=60$ with all $u_{ij}=+1$, and $J_{x}=1.0$, $J_{y}=-0.4$, $J_{z_{1}}=J_{z_{3}}=0.2$ for the $\mathcal{S}_{1}$ and $\mathcal{S}_{3}$ sections, and $J_{z_{2}}=3$ for the middle section $\mathcal{S}_{2}$. We can see that the MES $\gamma_{1}$ and $\gamma_{4}$ are respectively localized at the left and right ends of the ladder, while MES $\gamma_{2}$ and $\gamma_{3}$ are localized at the junctions between topological and nontopological sections of the ladder (the precise shape of MES can be understood more intuitively by mapping the spin ladder to two coupled $p$-wave superconducting wires; see Appendix \ref{sec:twowires}). The four zero-energy eigenvalues of $iA_{ij}^{u}$ (corresponding to $\gamma_{1,...
 ,4}$) reside inside a gap of about $1.7$ for this choice of parameters .
From Eq.~(\ref{eq:nu}) one concludes that it is possible to move from the topological to the nontopological phase by changing the relative strengths of $J_{x,y,z}$. Since MES will exist at the junction between sections in different topological phases, MES can be created, destroyed, and transported by locally (and adiabatically) changing the relative strengths of $J_{x,y,z}$ along the spin ladder. Finally, it is well-known that, in principle, exchange interactions can be controlled electrically (for atomistic or nano structures see, e.g., Refs.~\onlinecite{Wiesendanger2009,Hanson}). Thus, applying gates over portions of the spin ladder will allow one to move the MES along the ladder.
This is similar to what is done in superconducting wires with local tuning of the chemical potential, see Ref.~\onlinecite{AliceaWire2011}.

For the sake of completeness, we study in Appendix \ref{sec:correlations} long-range static correlations which exist in the topological phase only. In the standard honeycomb model spin spin correlations vanish rapidly with distance. \cite{Shankar2007,Nussinov2008} 
\color{black}

\section{Vortex-free (full) ground state}\label{sec:3}
\begin{figure}[h]
	\centering
		\includegraphics[width=0.4\textwidth]{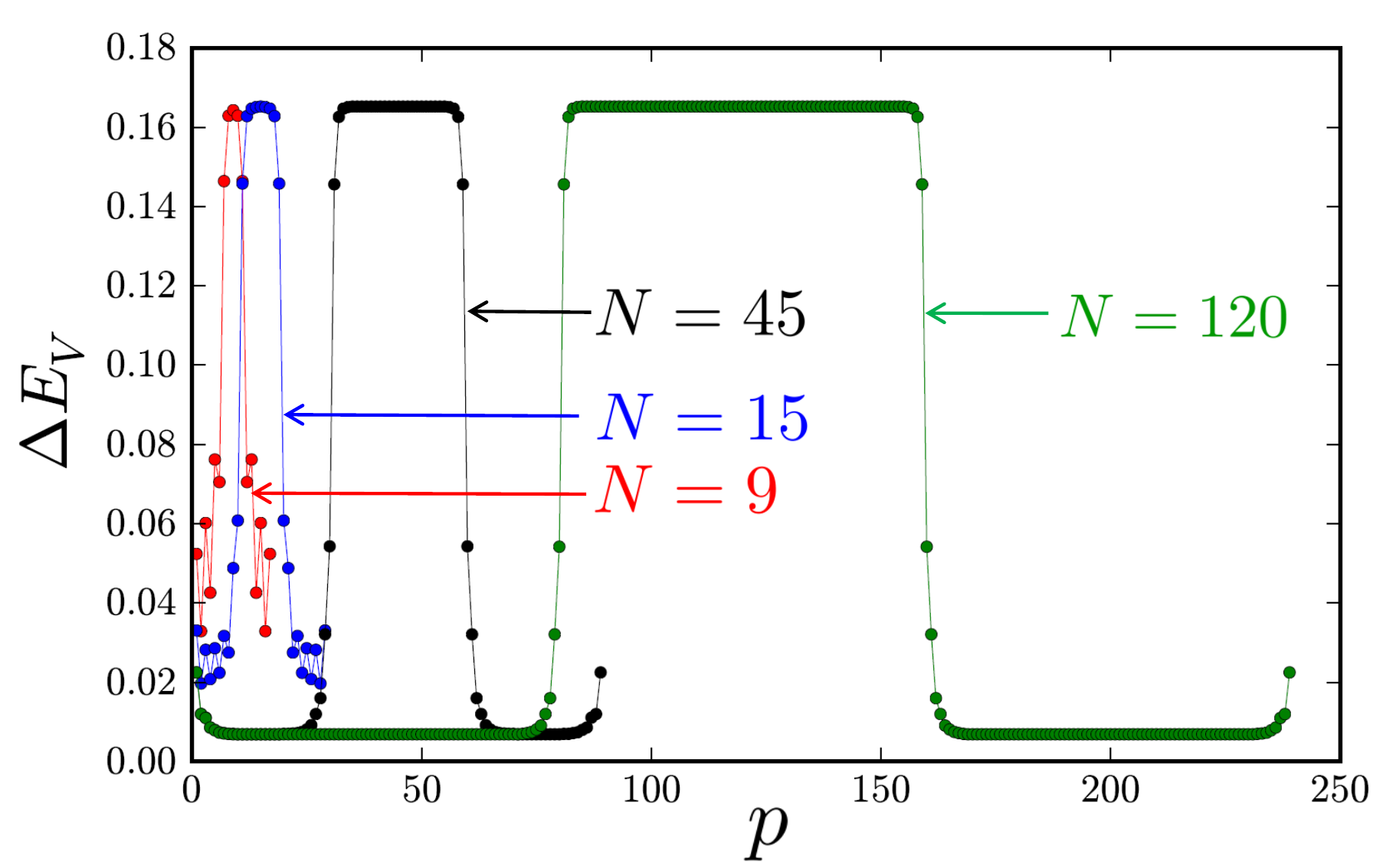}
	\caption{Excitation energy $\Delta E_{V}$ of a single vortex as a function of its position $p$ (square plaquette) on the ladder for different $N$. The values of $J_{x,y}$ and $J_{z_{ij}} $ are given in the main text. The junction between sections $\mathcal{S}_{1}$ and $\mathcal{S}_{2}$ is at $p=2N/3$ and between $\mathcal{S}_{2}$ and $\mathcal{S}_{3}$ at $p=4N/3$. Note the slight increase of $\Delta E_{V}$ (as compared to the bulk) for $N=45$ and $120$ when the vortex is located at the end of the ladder.}
	\label{fig:Vortex_Numerical}
\end{figure}

In this section we give numerical evidence that the ground state is either vortex-free or vortex-full for a certain range of $J_{x,y,z}$ parameters. 
It is tempting to invoke Lieb's theorem \cite{Lieb1994} (see also Ref.~\onlinecite{Macris1996}). However, this theorem is not directly applicable to our system since it requires periodic boundary conditions in the horizontal direction or $\vert J_{x}\vert=\vert J_{y}\vert$ with open boundaries (when the reflection plane is taken to be horizontal and going through the middle of the ladder).
However, different numerical checks lead us to conclude
that the ground state of our spin model is vortex-free/vortex-full for $\m{sgn}(J_{x})=(-/+)\m{sgn}(J_{y})$ and general $J_{z_{ij}}>0$. Figure \ref{fig:Vortex_Numerical} is a plot of the single-vortex energy for different $N$ and couplings $J_{x}=1.0$, $J_{y}=-0.4$, and $J_{z_{ij}}=J_{z_{1}}=J_{z_{3}}=0.2$ in sections $\mathcal{S}_{1}$ and $\mathcal{S}_{3}$, while $J_{z_{ij}}=J_{z_{2}}=4$ in section $\mathcal{S}_{2}$. The vortex energy converges quickly with $N$ and is positive. We also see that a vortex lying in the nontopological section $\mathcal{S}_{2}$ has a larger energy since $J_{z}$ is stronger there. Figure~\ref{fig:Vortex_Numerical} reveals that the energy becomes independent of the vortex position in the bulk of the ladder when $N$ is large. However, the ground state is vortex-free or vortex-full and such additional degeneracies are not present. All ground-state degeneracies can thus be associated to MES.\color{black}

We have numerically investigated the energy of multivortex configurations and found that, although the vortex-vortex interaction is attractive, the attraction is not strong enough to favor the creation of additional vortices and the ground state remains a vortex-free state (see Appendix \ref{sec:Vortex} for more details). Additional numerical checks with different $J_{x,y,z}$ configurations are also reported there. For all the numerical checks we performed, the conclusion remains the same: The ground state is vortex-free. Since changing the sign of $J_{x,y}$ is equivalent to $u_{ij}^{x,y}=-1$ for the corresponding bond, the system with $\m{sgn}(J_{x})=\m{sgn}(J_{y})$ has thus a vortex-full ground state, as expected.
Although an analytical proof, to the best of our knowledge, is lacking, we conjecture that Lieb's theorem can be formally extended to the spin ladder considered in this work. 
\section{Robustness of the topological degeneracy}\label{sec:4}
As mentioned in the Introduction, topological order does not exist in one-dimensional systems. \cite{Verstraete2005,Wen2011,Schuch2011} Especially, if no symmetry constraints are present, it is always possible to find local perturbations which split the Majorana degeneracy.

In line with this general result, we are thus interested in determining what type of local spin perturbations can split the topological degeneracy and against which type of perturbations the MES of the inhomogeneous ladder are robust. Indeed, as we shall see, only certain forms of local perturbations can split the topological degeneracy. A symmetry-based analysis of perturbations that can split the topological degeneracy into different spin ladders has recently been performed in Ref.~\onlinecite{Wen2012}. 

Here we demonstrate that single-body terms cannot fully split the topological degeneracy present in the inhomogeneous ladder and that at least two-body perturbations are required. Furthermore, we also demonstrate that a twofold degeneracy always remains when $x$, $y$, or $z$ inhomogeneous magnetic fields are applied. The presence of inhomogeneities in the ladder is important, since much simpler perturbations destroy the topological degeneracy of the homogeneous ladder as we show below.
\subsection{Homogeneous ladder}
Here we first focus on the homogeneous ladder, while we study the inhomogeneous ladder in the next section. In order to study the robustness of MES, we recall that the spin system carries four additional $b$ Majoranas arising from the $b$ operators at the ends of the ladder ($b_{1}^x$, $b_{2}^y$, $b_{4N-1}^y$, and $b_{4N}^x$) which are completely decoupled from the Hamiltonian $\widetilde{H}_{u}$. These $b$ Majoranas are always present in the model (in the extended space) independent of the strength of the couplings $J_{x,y,z}$. Such a model carries six Majoranas in the topological phase, namely two spatially separated $c$ Majoranas and two $b$ Majoranas at each end. From Eq.~(\ref{eq:mapping}) it is thus clear that a local single-body perturbation $V=h_{1}\sigma_{1}^{x}+h_{4N}\sigma_{4N}^{x}$ will combine two of the three Majoranas at each end of the ladder, leaving only one Majorana at the right and one Majorana at the left end of the ladder (and thus one zero-energy fermion state). It has recently been shown \cite{Chesi2011} that, given a certain vortex configuration and lattice topology, only half of the states in the extended space are physical and these physical states have a definite fermionic parity. If the physical states have even parity, then the remaining zero-energy fermion state is unfilled, while it is filled for physical states with odd parity. In any case this means that one of the states of the remaining zero-energy fermion is unphysical. We thus conclude that the topological degeneracy of the homogeneous ladder has been lifted by $x$ single-body terms only.
\subsection{Inhomogeneous ladder}
Let us now consider the inhomogeneous ladder. The ladder possesses eight MES, namely four spatially separated $\gamma_{i}$ (see Fig.~\ref{fig:Chain_5}) and two $b$ Majorana operators at each end. An argument identical to the one presented for the homogeneous ladder is thus applicable here and single-body local perturbations  $V=h_{1}\sigma_{1}^{x}+h_{4N}\sigma_{4N}^{x}$ will mix one $b$ with one $c$ Majorana at each end of the ladder. Since four Majoranas are left in the extended space, we still have a twofold degeneracy in the physical space. The corresponding states in a given $u$ configuration are denoted by $\vert\Psi_{m}^u\rangle$ with $m=1,2$. The four Majoranas are well-separated from each other; hence, it seems impossible to remove this remaining degeneracy with local perturbations $V$; indeed $\langle\Psi_1^u\vert V\vert\Psi_2^u\rangle=0$. However, one has to be careful with the fact that the physical states are given by symmetrization over all gauge operators, see Eq.~(\ref{eq:symmetrization}). The physical vortex-free degenerate ground states $\vert\Psi_{1,2}^{\m{VF}}\rangle$ are thus given by
\begin{equation}\label{eq:projec}
\vert\Psi_{1,2}^{\m{VF}}\rangle=\underbrace{\prod_{i}\left(\frac{1+D_i}{2}\right)}_{\mathcal{P}}\vert\Psi_{1,2}^u\rangle,
\end{equation}
where $u$ is the configuration with all $u_{ij}=+1$.

From Eq.~(\ref{eq:projec}), we realize that the projection onto the physical subspace introduces a string of $D_i$ operators and that the spin to fermion mapping (\ref{eq:mapping}) is only apparently local. Similar to the case encountered with a Jordan-Wigner transformation, \cite{Tserkovnyak2011} the nonlocality of the mapping means that a local spin perturbation is nonlocal in the fermionic language and can thus connect well-separated MES. In order to check whether a local perturbation $V$ splits the Majorana degeneracy, we thus need to calculate the following matrix elements:
\begin{equation}
\langle\Psi_{m}^{\m{VF}}\vert V\vert\Psi_{n}^{\m{VF}}\rangle=\langle\Psi_{m}^u\vert V\mathcal{P}\vert\Psi_{n}^u\rangle,
\end{equation}
with $m,n\in\{1,2\}$. Since $\mathcal{P}$ contains string of $D_{i}$ operators, we need to check whether $V$ introduces transition between $\vert\Psi_{1,2}^u\rangle$ and $\prod_{i\in\Lambda}D_i\vert\Psi_{1,2}^u\rangle$, where $\Lambda$ is a subset of indices. If this is the case, then the perturbation $V$ can split the remaining degeneracy.

In the following we explicitly construct a local perturbation which splits the remaining degeneracy in agreement with the general result of Refs.~\onlinecite{Verstraete2005,Wen2011,Schuch2011} stating that topological order is impossible in one-dimensional systems. Furthermore, we show that such a perturbation has to contain at least two-body terms.

A necessary criterion for a local perturbation $V$ to split the degeneracy is that it does not create vortices. This is so because the degenerate ground states are vortex-free and states with different $u$ configurations are orthogonal. Let us thus first consider all possible single-body terms which do not create any vortex and show that they cannot split the degeneracy. The effect of $V_{1}=\epsilon_1\sigma_{1}^{x}$ and $V_{4N}=\epsilon_{4N}\sigma_{4N}^{x}$ has been considered already. The remaining single-body terms which do not create vortices are
\begin{eqnarray}
&V_2=\epsilon_2\sigma_2^y,\,\,\,& V_{4N-1}=\epsilon_{4N-1}\sigma_{4N-1}^y.
\end{eqnarray}
For these single-body perturbations, the only string of $D_i$ operators which leaves the $u$-configuration invariant is the product of all $D_i$, i.e., $\prod_{i=1}^{4N}D_i$. As an explicit example, let us now consider the effect of $V_{2}$:
\begin{eqnarray}
\sigma_{2}^y\prod_{i}^{4N}D_{i}&=&i b_{2}^y c_2 \prod_{i=1}^{4N}b_{i}^xb_{i}^yb_{i}^zc_{i}\\
&\propto& ib_{2}^yc_2 b_{1}^xb_{2}^yb_{4N-1}^yb_{4N}^x\prod_{\langle ij\rangle}u_{ij}\prod_{i=1}^{4N}c_i\\
&\propto& ib_{2}^yc_2\hat{\pi},
\end{eqnarray}
where $\hat{\pi}=b_{1}^xb_{2}^yb_{4N-1}^yb_{4N}^x\prod_{i=1}^{4N}c_{i}$ is a parity operator and we used the fact that all $u_{ij}=+1$. We thus have
\begin{eqnarray}
\langle\Psi_{m}^{u}\vert ib_{2}^{y}c_{2}\hat{\pi}\vert\Psi_{n}^{u}\rangle=\pi\langle\Psi_{m}^{u}\vert ib_{2}^{y}c_{2}\vert\Psi_{n}^{u}\rangle=0,
\end{eqnarray}
where $m,n\in\{1,2\}$, $\pi$ is the parity of $\vert\Psi_{1,2}^{u}\rangle$ and the last equality comes from the fact that $c_{2}$ creates a finite-energy fermion (the contribution of $c_{2}$ in the remaining MES is exponentially supressed with system's size). We thus finally conclude that single-body perturbations cannot fully split the topological degeneracy.

In the following we show that two-body interactions are enough to destroy the remaining MES and we construct explicitly a perturbation which splits them:
\begin{equation}\label{eq:twobodyper}
V=\epsilon\sigma_{M-1}^x\sigma_M^{y},
\end{equation}
where $M$ is even and belongs to the nontopological section $\mathcal{S}_2$, see Fig.~(\ref{fig:Chain_5}). An important point to notice is that such a perturbation will not create any vortices since $\sigma_{M-1}^x$ changes the values of $u_{M-2,M-1}^x$ while $\sigma_M^{y}$ changes the value of $u_{M-3,M}^y$. If this was not the case, then all matrix elements $\langle\Psi_{m}^{u}\vert V\prod_{i\in\Lambda}D_i\vert\Psi_{n}^u\rangle$ would trivially vanish, since states with different $u$ configuration are orthogonal. 
The fact that states with different $u$ configurations are orthogonal implies that the matrix elements $\langle\Psi_{m}^u\vert\prod_{i\in\Lambda}D_i\sigma_{M-1}^x\sigma_M^{y}\vert\Psi_{n}^u\rangle$ can be different than zero only if there exists a string of $D_i$ operators which connect the two different configurations of $u$. Having in mind that $D_i$ changes the value of $u$ for the three links connected to site $i$, it is easy to check that $\prod_{i=1}^{M-2}D_{i}$ introduces transitions between these two configurations of $u$. We have
\begin{eqnarray}\label{eq:D}
\prod_{i=1}^{M-2}D_i\propto b_{1}^{x}b_{2}^{y}b_{M-3}^{y}b_{M-2}^{x}\prod_{i=1}^{M-2}c_{i},
\end{eqnarray}
where the other $b$ operators combine into $u$ operators which are all equal to $1$. From Eq.~(\ref{eq:D}) we then obtain
\begin{eqnarray}\label{eq:D2}
\sigma_{M-1}^x\sigma_{M}^{y}\prod_{i=1}^{M-2}D_i\propto b_{2}^{y}b_{1}^{x}\prod_{i=1}^{M}c_{i}
\end{eqnarray}
The string of $c_{i}$ operators in Eq.~(\ref{eq:D2}) is nonlocal and connect MES $b_{2}^{y}$ with MES $c_{M}$. This makes it possible to conclude that $\sigma_{M-1}^x\sigma_{M}^{y}\prod_{i=1}^{M-2}D_i$ can split the remaining topological degeneracy. However, checking that the matrix elements are really different from zero, i.e.,
\begin{eqnarray}
\langle\Psi_{m}^u\vert\prod_{i=1}^{M-2}D_i\sigma_{M-1}^x\sigma_M^{y}\vert\Psi_{n}^u\rangle\neq 0,
\end{eqnarray}
requires determining whether the string operator $\prod_{i=1}^{M}c_{i}$ makes it possible to come back to the ground subspace. This is, in principle, the case since $c_{i}$'s are superpositions of all eigenmodes. In Appendix \ref{sec:newmapping} we use a different approach and show explicilty that $V$ can indeed induce splitting between the Majorana states (all these considerations remain valid for a perturbation $V=\epsilon\sigma_{M}^{y}\sigma_{M+1}^x$ with odd $M$ as well). 
Although single-body terms are not able to fully split the topological degeneracy, we showed that local two-body terms are. From this result, it appears that MES are quite fragile against external magnetic fields since two-body interactions are easily generated in second-order perturbation theory. However, on the positive side, we show below that the topological degeneracy is protected against inhomogeneous $x$, $y$, and $z$ magnetic field components.

\subsection{Protection against magnetic fields aligned along $x$, $y$, or $z$ direction}

Let us consider a local perturbation $V$ which represents an inhomogeneous magnetic field in direction $\alpha=x,y,z$:
\begin{equation}
V=\sum_{i}h_{i}\sigma_{i}^{\alpha}.
\end{equation}
As we demonstrated in the previous section, single-body terms cannot fully split the topological degeneracy. As a consequence, the perturbation $V$ is not able to split the topological degeneracy at first order. We thus ask here the question whether this is possible at higher orders. The answer, in principle, is yes. For MES well-separated by $L$ sites, the degeneracy can obviously be split in $L^{\m{th}}$ order. However, $L^{\m{th}}$-order terms are exponentially suppressed with system's size and the effect of such perturbation is thus negligible for large systems. We are then interested in determining whether this is possible at order $n$ independent of system's size.

As mentioned previously, a necessary condition for a local perturbation to split the degeneracy is that it does not create any vortices. The good candidates $V^n$ generated at order $n$ have thus to be products of terms $\sigma_{i}^{\alpha}\sigma_{j}^{\alpha}$ and $\sigma_{1,4N}^{x}$ or $\sigma_{2,4N-1}^{y}$, where $i,j$ are nearest-neighbor sites along an $\alpha$-link $\langle i,j\rangle_{\alpha}$. Note that $V^n$ of this form not only do not create vortices, but also do not change the configuration of $u$. Thus, the two matrix elements which need to be considered are
\begin{equation}
\langle\Psi_{m}^{u}\vert V^n\vert\Psi_{n}^{u}\rangle\,\,\,\,\,\m{and}\,\,\,\,\,\langle\Psi_{m}^{u}\vert V^n\prod_{i=1}^{4N}D_{i}\vert\Psi_{n}^{u}\rangle,
\end{equation}
with $m,n\in\{1,2\}$, since $\prod_{i=1}^{4N}D_{i}$ leaves the configuration of $u$ invariant. If we recall that (the remaining four) MES are well-separated from each other, it is then clear that $\langle\Psi_{m}^{u}\vert V^{n}\vert\Psi_{n}^{u}\rangle=0$ for $n<L$. Furthermore, as mentioned previously, the operator $\prod_{i=1}^{4N}D_{i}$ is proportional to the parity operator of the eigenmodes \cite{Chesi2011} and we thus have for $n<L$
\begin{eqnarray}
\langle\Psi_{m}^{u}\vert V^n\prod_{i=1}^{4N}D_{i}\vert\Psi_{n}^{u}\rangle\propto \pi\langle\Psi_{m}^{u}\vert V^n\vert\Psi_{n}^{u}\rangle=0,
\end{eqnarray}
where $\pi$ is the eigenmodes parity of $\vert\Psi_{1,2}^{u}\rangle$.

From this we conclude that the effect of $x$, $y$, or $z$ magnetic field components on the inhomogeneous spin ladder is exponentially suppressed with system size. In other words, the topological degeneracy cannot be fully split by inhomogeneous magnetic fields purely along the $x$, $y$, or $z$ direction.

\section{Braiding MES in a trijunction setup}\label{sec:5}
\begin{figure}[h]
	\centering
		\includegraphics[width=0.45\textwidth]{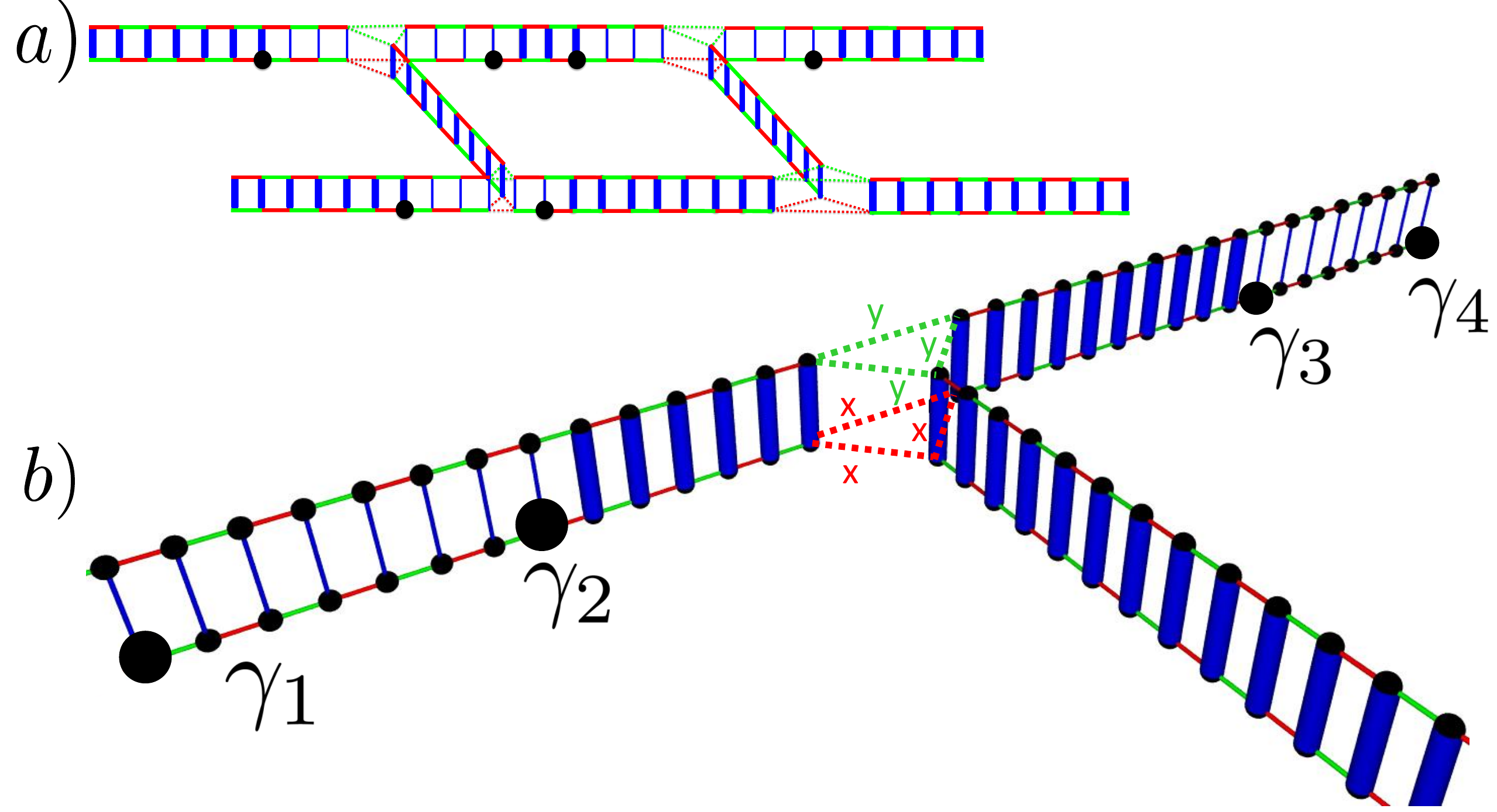}
	\caption{Network of spin ladders. a) Majoranas are exchanged through the trijunctions, which are shown in detail in $b$). The connection between the three spin ladders is given by Ising couplings $J_{x}$ (red dashed lines) and $J_{y}$ (green dashed lines). Braiding is performed by varying $J_{z}$, see main text.
	MES $\gamma_{1,...,4}$ (large dots) are localized at the left and right ends of the ladder and at the junction between topological (thin $z$-links) and nontopological (thick $z$-links) sections.}
	\label{fig:Trijunction}
\end{figure}

The recent demonstration of non-Abelian character of MES in $p$-wave wires \cite{AliceaWire2011} makes the possibility to realize topological quantum computing by braiding MES an attractive and promising method. Majorana end states in spin ladders can be moved by locally tuning the value of $J_{z}$ couplings (similar to the local tuning of chemical potential in $p$-wave wires) and it is thus in principle also possible to braid them (following the scheme of Ref.~\onlinecite{AliceaWire2011}) in the trijunction setup presented in Fig.~\ref{fig:Trijunction}. The trijunction in Fig.~\ref{fig:Trijunction} possesses the interesting property that no spurious Majorana modes are created in the course of braiding. Indeed, when all three parts building the trijunction are in the topological phase, the $J_{x,y}$ couplings (dashed lines in Fig. 5) between the MES will combine two of them into an ordinary complex fermion. We note that such braiding processes in the simpler $xx-yy$ chain would be difficult to realize because of the many additional zero-energy modes, see Appendix \ref{sec:proliferation}. This is why we are interested in inhomogeneous spin ladders where such additional degeneracies do not appear and all Majorana modes are well-separated from each other.

In order to be of use for topological quantum computing, an important property of MES is that they follow non-Abelian exchange statistics (as it is the case in $p$-wave wires \cite{AliceaWire2011}). As mentioned above, the projection operator (onto the physical subspace) $\mathcal{P}$ introduces strings of operators and renders mapping (\ref{eq:mapping}) only apparently local. In such a case the study of braiding statistics is more complicated and especially mapping the Hamiltonian onto Kitaev's toy model Hamiltonian is not enough to conclude anything about the statistics of MES. \cite{Tserkovnyak2011} We leave here the question of the MES statistics as an open problem. 

To perform a robust topological quantum computation by exchanging MES, the two-body perturbations splitting the topological degeneracy have to be excluded. This requirement is analogous to the parity conservation requirement in $p$-wave wires and can, in principle, be achieved by carefully screening magnetic fields. Since Ising interactions can be tuned by means of electric fields only,\cite{Wiesendanger2009,Hanson} magnetic fields are not required for braiding and there is no intrinsic contradiction between exchanging MES and screening external magnetic fields. 

\color{black}


\section{Conclusions}\label{sec:6}
We have proposed inhomogeneous spin ladders and shown that they support a topological phase with localized Majorana states. We have studied the robustness of MES under local perturbations and demonstrated that single-body perturbations are not enough to split the topological degeneracy. We have explicitly constructed a two-body perturbation which does this, in agreement with the general fact that topological order is not possible in one-dimensional systems. \cite{Verstraete2005,Wen2011,Schuch2011} On the positive side, we showed that the topological degeneracy cannot be destroyed by inhomogeneous magnetic fields aligned purely along $x$, $y$, or $z$ direction. Finally, we have presented a trijunction setup where MES can be exchanged similar to Ref.~\onlinecite{AliceaWire2011}.

While the spin ladders envisioned here are not yet available, we hope that the present study provides a strong encouragement for their experimental realization since they represent a promising framework for the realization and manipulation of MES.


\section{Acknowledgement}
We thank Xiao-Gang Wen, Xie Chen, Bei Zeng, Luka Trifunovic, and Diego Rainis for useful discussions. We acknowledge support from the Swiss NF, NCCRs Nanoscience and QSIT, and SOLID. SC acknowledges financial support from NSERC, CIFAR, FQRNT, and INTRIQ.
\color{black}

\appendix
\section{Mapping to two coupled Kitaev $p$-wave superconducting wires}\label{sec:twowires}
\begin{figure}[h]
	\centering
		\includegraphics[width=0.45\textwidth]{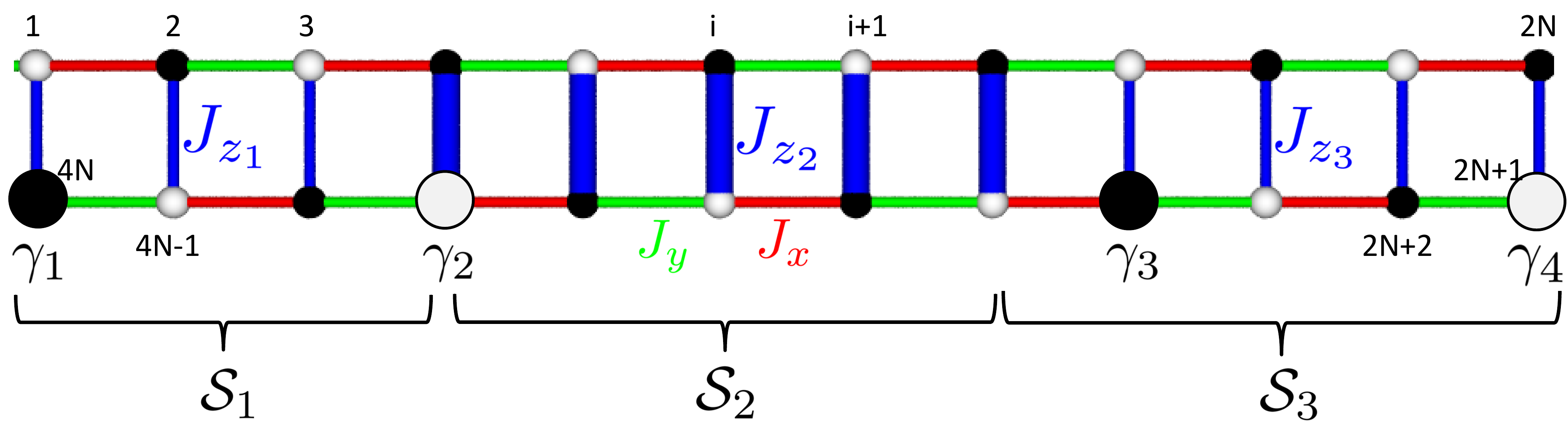}
	\caption{Inhomogeneous Kitaev spin ladder. This spin ladder possesses two topological sections $\mathcal{S}_{1}$ and $\mathcal{S}_{3}$ (thin $z$ links with couplings $J_{z_{1}}$ and $J_{z_{3}}$ which we choose to be equal, i.e., $J_{z_{1}}=J_{z_{3}}=J_{z}$) separated by a nontopological section $\mathcal{S}_{2}$ (thick $z$ links with couplings $J_{z_{2}}=J_{z^{\prime}}$). The main components of the four MES wavefunctions $\gamma_{1,...,4}$ lie on the lower sites for $J_{x}>J_{y}$ and are represented by large dots.}
	\label{fig:Chain_5_Supp}
\end{figure}
As presented in the main text, the model we consider possesses three different sections $\mathcal{S}_{1}$, $\mathcal{S}_{2}$, and $\mathcal{S}_{3}$ which are distinguished by the value of the $J_{z_{ij}}$ couplings [see Fig.~\ref{fig:Chain_5_Supp}]. We focus on the vortex-free and vortex-full sectors where we choose $J_{x,y,z,z'}$ such that $\mathcal{S}_{1}$ and $\mathcal{S}_{3}$ are topological, while section $\mathcal{S}_{2}$ is nontopological. This system carries four MES: $\gamma_{1}$ and $\gamma_{4}$ at  the left and right end of the ladder, respectively, while $\gamma_{2}$ and $\gamma_{3}$ sit at the junction between topological ($\mathcal{S}_{2}$) and nontopological ($\mathcal{S}_{1,3}$) sections of the ladder. 

Let us focus on a topological section, say $\mathcal{S}_{1}$, in the vortex-free sector (i.e., $u_{ij}=+1$),  and study the location of MES $\gamma_{1,2}$ and their behavior under the modification of $J_{x,y,z}$ couplings. We disregard here the presence of the two other sections $\mathcal{S}_{2,3}$. It is useful to consider our spin system as two $xx$-$yy$ chains coupled via $J_{z}$ Ising couplings. Let us now introduce the following complex fermion operators:
\begin{equation}
d_{j}=\frac{1}{2}(c_{2j-1}+ic_{2j})\,\,\,\,\m{and}\,\,\,\,d_{j}^{\dagger}=\frac{1}{2}(c_{2j-1}-ic_{2j}),
\end{equation}
with $j=1...2N$ (the site labeling is shown in Fig.~\ref{fig:Majorana_S_W}), $\{d_{j},d_{j'}\}=0$, and $\{d_{j},d_{j'}^{\dagger}\}=\delta_{jj'}$. Then the upper ($u$) $xx$-$yy$ chain is mapped to the Kitaev model for a one-dimensional $p$-wave superconductor,  \cite{LiebSchultzMattis,Kitaev2001,DamleHuse2001}
\begin{equation}
\label{upperwire}
H^{u}=-\mu^{u}\sum_{j=1}^{N}d_{j}^{\dagger}d_{j}-\sum_{j=1}^{N-1}\left(t^{u}d_{j}^{\dagger}d_{j+1}+\Delta ^{u} d_{j}d_{j+1}+\m{h.c.}\right),
\end{equation}
with $\mu^{u}=2J_{x}$ and $t^{u}=-\Delta^{u}=J_{y}$, while the lower ($l$) $xx$-$yy$ spin chain is mapped to
\begin{equation}
\label{lowerwire}
H^{l}=-\mu^{l}\sum_{j=N+1}^{2N}d_{j}^{\dagger}d_{j}-\sum_{j=N+1}^{2N-1}\left(t^{l}d_{j}^{\dagger}d_{j+1}+\Delta ^{l}d_{j}d_{j+1}+\m{h.c.}\right),
\end{equation}
with $\mu^{l}=2J_{y}$ and $t^{l}=-\Delta^{l}=J_{x}$.

The $J_{z}$ spin couplings between upper and lower $xx$-$yy$ chain leads to a hopping term $H^{ul}$ between upper and lower wire in the fermionic representation,
\begin{equation}
H^{ul}=-\sum_{j=1}^{N}\left(t^{ul}d_{j}^{\dagger}d_{2N-(j-1)}+\m{h.c.}\right),
\end{equation}
where $t^{ul}=2J_{z}$.

Let us first focus on the case $J_{z}=0$. Then, the system consists of two decoupled wires, Eqs. (\ref{upperwire}) and (\ref{lowerwire}), and we can distinguish between the following cases: If $J_{x}>J_{y}$, then the upper wire lies in the nontopological and the lower wire in the topological phase, i.e., the two MES are localized in the lower wire, one at the left and one at the right end; vice versa for $J_{y}>J_{x}$.

When the $z$-couplings are turned on, i.e. $J_{z}>0$, then the MES spread over both the upper and the lower wires as shown in Fig.~3 of the main text. If $J_{z}$ increases, then the MES continue to spread until they completely split when $\vert J_{z}\vert>\vert J_{x}+J_{y}\vert$ in the vortex-free sector and $\vert J_{z}\vert >\vert J_{x}-J_{y}\vert$ in the vortex-full sector, see Eq.~(6) in the main text. It is also straightforward to understand the exact site localization of the two MES. For $J_{x}>J_{y}$, most of the weight of the left $\gamma_{1}$ (right $\gamma_{2}$) MES resides at respectively the first and last site of the lower $xx$-$yy$ chain and spreads only over $A$ ($B$) sublattice sites. Indeed, the $J_{x,y,z}$-couplings between spins residing on different sublattices forbids $\gamma_{1}$ ($\gamma_{2}$) to spread over $B$ ($A$) sites. Similarly, for $J_{y}>J_{x}$, most of the weight of the left $\gamma_{1}$ (right $\gamma_{2}$) MES resides at, respectively, the first and last sites of the upper $xx$-$yy$ chain and spreads only over $B$ ($A$) sublattice sites.
\section{Vortex-free and vortex-full ground states}\label{sec:Vortex}
\begin{figure}[h]
	\centering
		\includegraphics[width=0.47\textwidth]{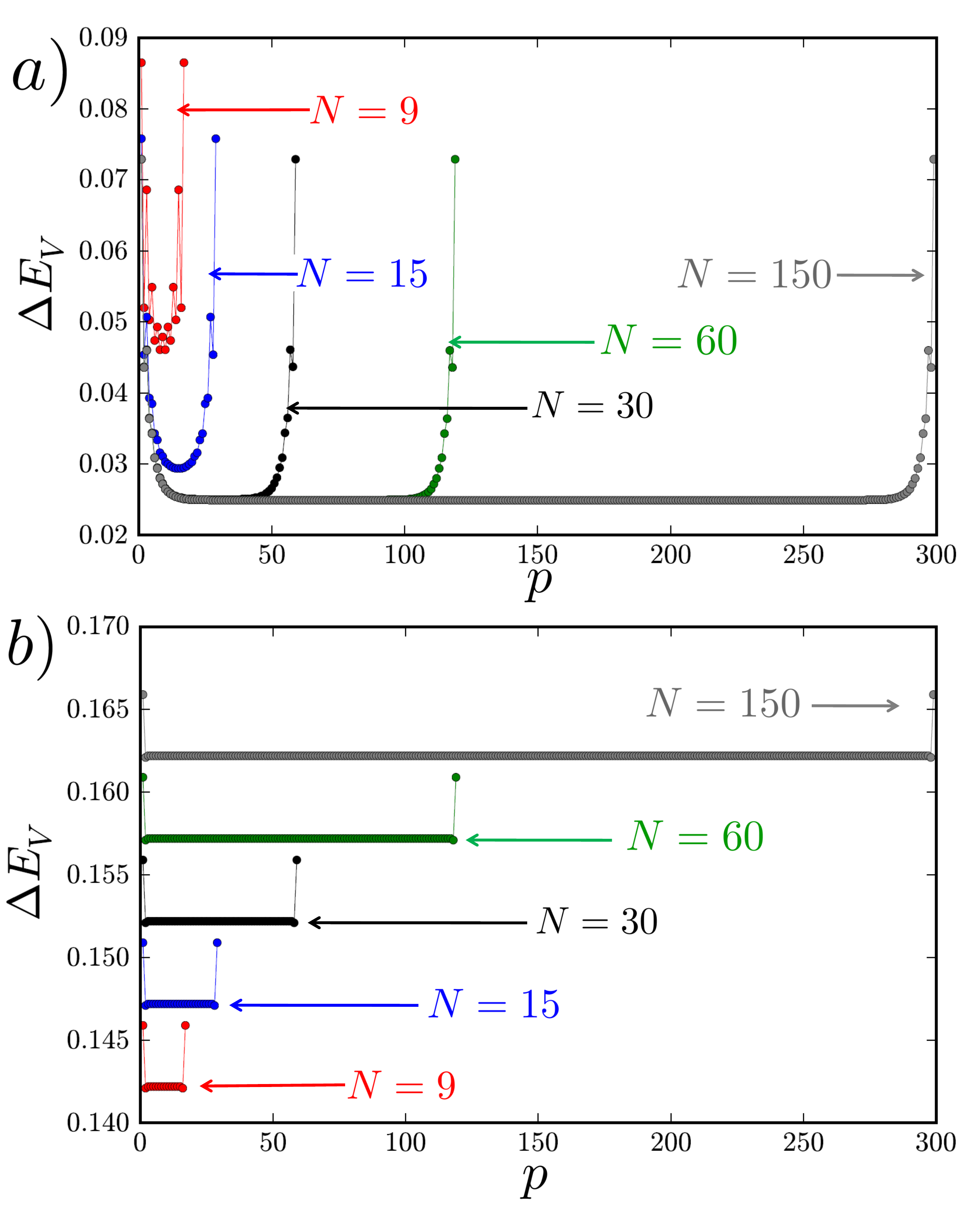}
	\caption{Energy $\Delta E_{V}$ of a single vortex as a function of its position $p$ on the ladder. We recall that a vortex can be placed at $2N-1$ different positions on a ladder with $N$ unit cells. The five different curves correspond to $N=9,15,30,60,150$. We see a clear difference between the vortex energy in the bulk and near the boundaries: Boundary effects increase the energy of a vortex lying near to one end of the ladder. It is also worth pointing out that the vortex energy converges quickly (with $N$) to its thermodynamic limit value. We see that the vortex energy is positive for each curve irrespective of the vortex's position. This plot thus supports our claim that the ground state is vortex free. The values of the different couplings are: $a)$ $J_{x}=1.0$, $J_{y}=-0.5$, $J_{z}=0.3$ in $\mathcal{S}_{1,3}$, and $J_{z'}=0.3$ in $\mathcal{S}_{2}$. $b)$ $J_{x}=1.0$, $J_{y}=-0.65$, $J_{z}=4.3$ in $\mathcal{S}_{1,3}$ and $J_{z'}=4.3$ in $\mathcal{S}_{2}$
 . The curves for $N=15,30,60,150$ are shifted vertically by $0.005, 0.01, 0.015$, and $0.02$, respectively, for clarity.}
	\label{fig:Vortex_energy_supplement_1}
\end{figure}

\begin{figure}[h]
	\includegraphics[width=0.47\textwidth]{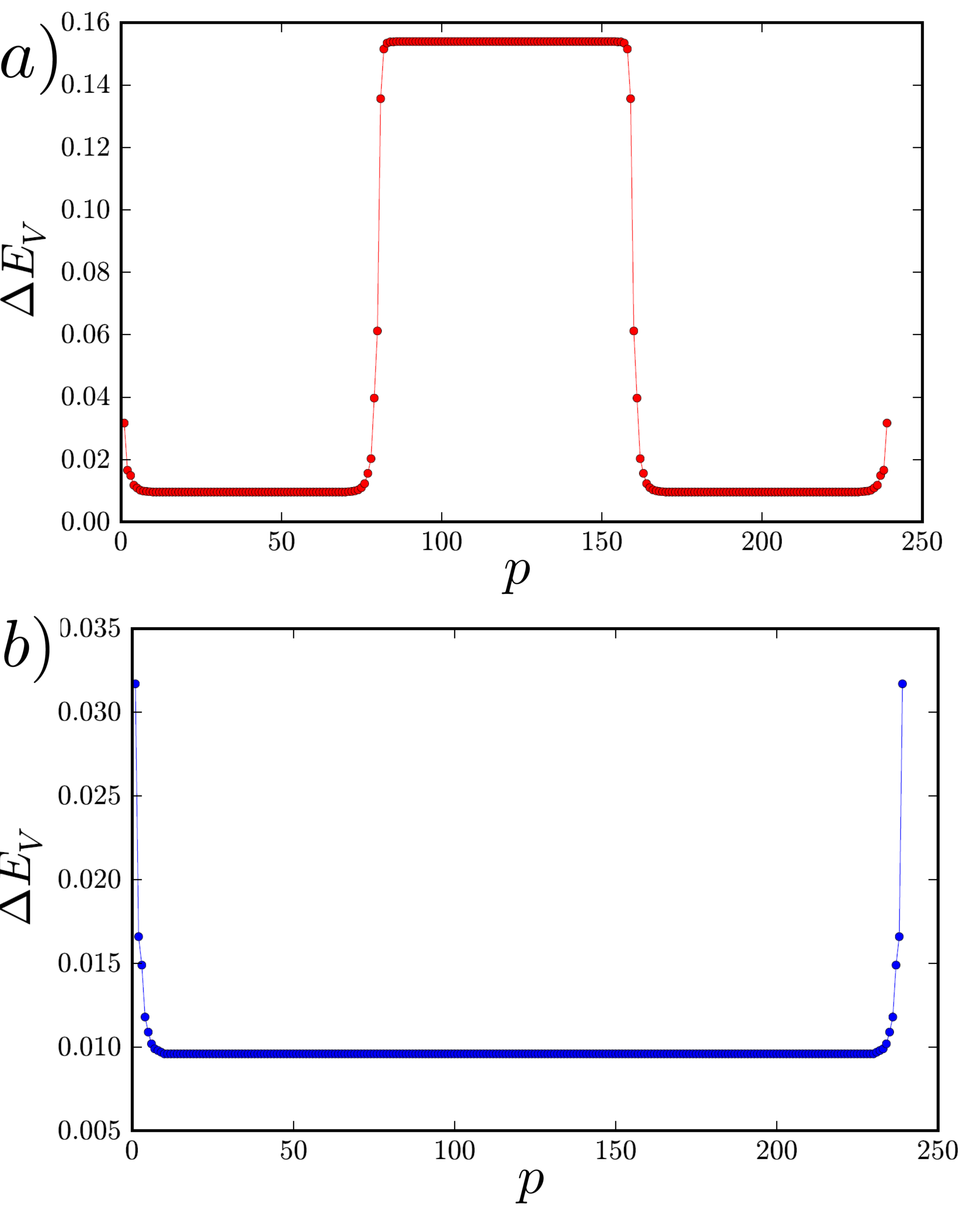}
	\caption{$a,b)$ Energy $\Delta E_{V}$ of a single vortex as a function of its position $p$ on a ladder with $N=120$, $J_{x}=1.0$, $J_{y}=-0.37$, $J_{z}=0.25$ in $\mathcal{S}_{1,3}$, while $J_{z'}=4$ in $\mathcal{S}_{2}$ for $a)$ and $J_{x}=1.0$, $J_{y}=-0.37$, $J_{z}=0.25$ in $\mathcal{S}_{1,3}$ and $J_{z'}=0.25$ in $\mathcal{S}_{2}$ for $b)$. The junction between sections $\mathcal{S}_{1}$ and $\mathcal{S}_{2}$ is at $p=2N/3$ and between $\mathcal{S}_{2}$ and $\mathcal{S}_{3}$ is at $p=4N/3$}
	\label{fig:Vortex_energy_supplement_4_1}
\end{figure}

\begin{figure}[h]
	\includegraphics[width=0.47\textwidth]{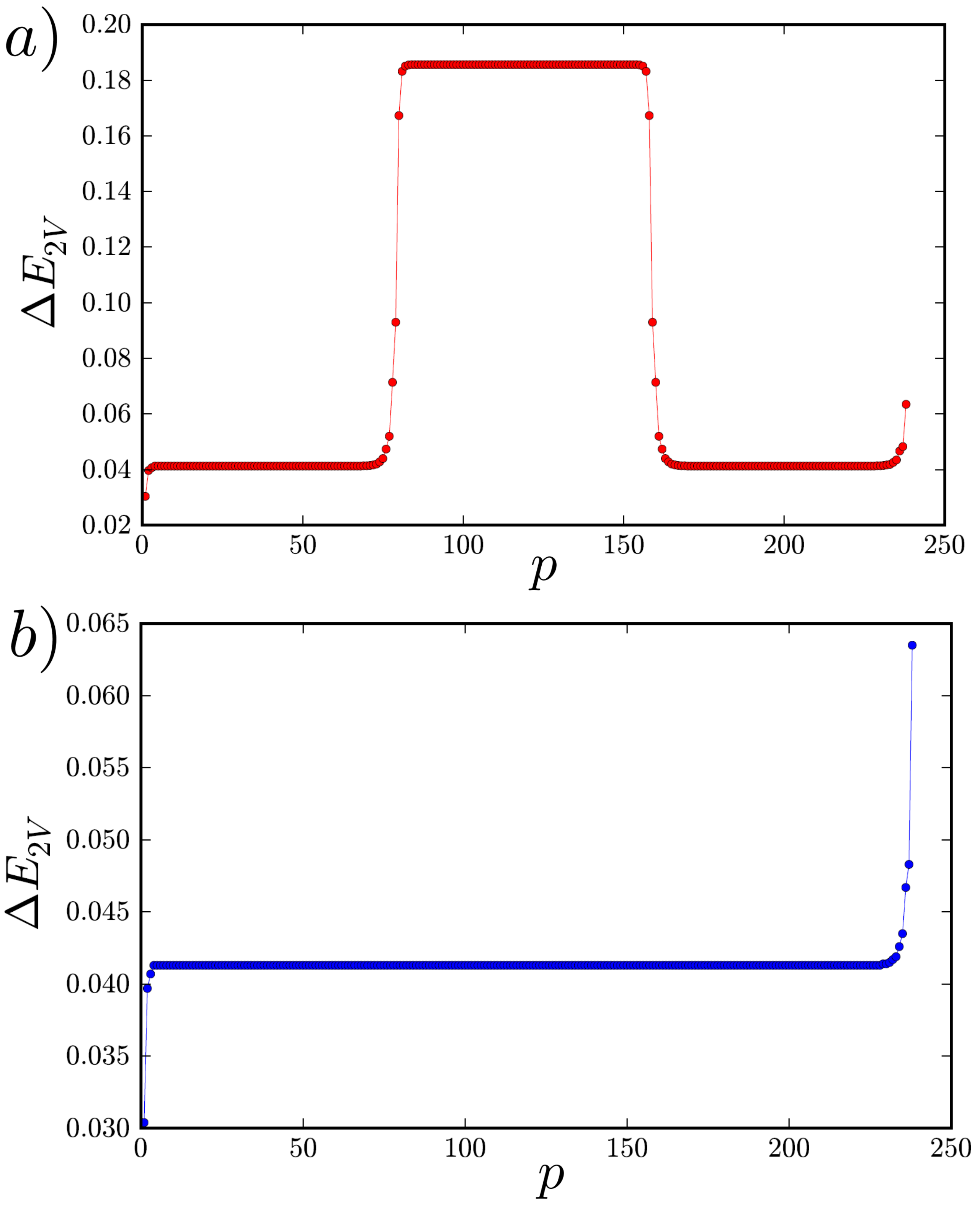}
	\caption{ $a,b)$ Energy $\Delta E_{2V}$ of two vortices as a function of the position $p$ of the second vortex. The first vortex lies on the $p=1$ square plaquette. The $J_{x,y,z,z'}$ parameters are chosen respectively as in Fig.~\ref{fig:Vortex_energy_supplement_4_1}. The vortex-vortex interaction is attractive and rapidly decaying as a function of distance between the two vortices. Indeed, already for $p=5$ the energy of the two vortices is roughly $0.04$ [$0.03$ (energy of the vortex at the boundary $p=1$) plus $0.01$ (energy of the vortex in the bulk)]. However, as mentioned in the main text, the attraction is never strong enough to favor the creation of vortices and the 
 energy of the two vortices is always positive. The junction between sections $\mathcal{S}_{1}$ and $\mathcal{S}_{2}$ is at $p=2N/3$ and between $\mathcal{S}_{2}$ and $\mathcal{S}_{3}$ is at $p=4N/3$}
	\label{fig:Vortex_energy_supplement_4_2}
\end{figure}

\begin{figure}[h]
	\centering
		\includegraphics[width=0.47\textwidth]{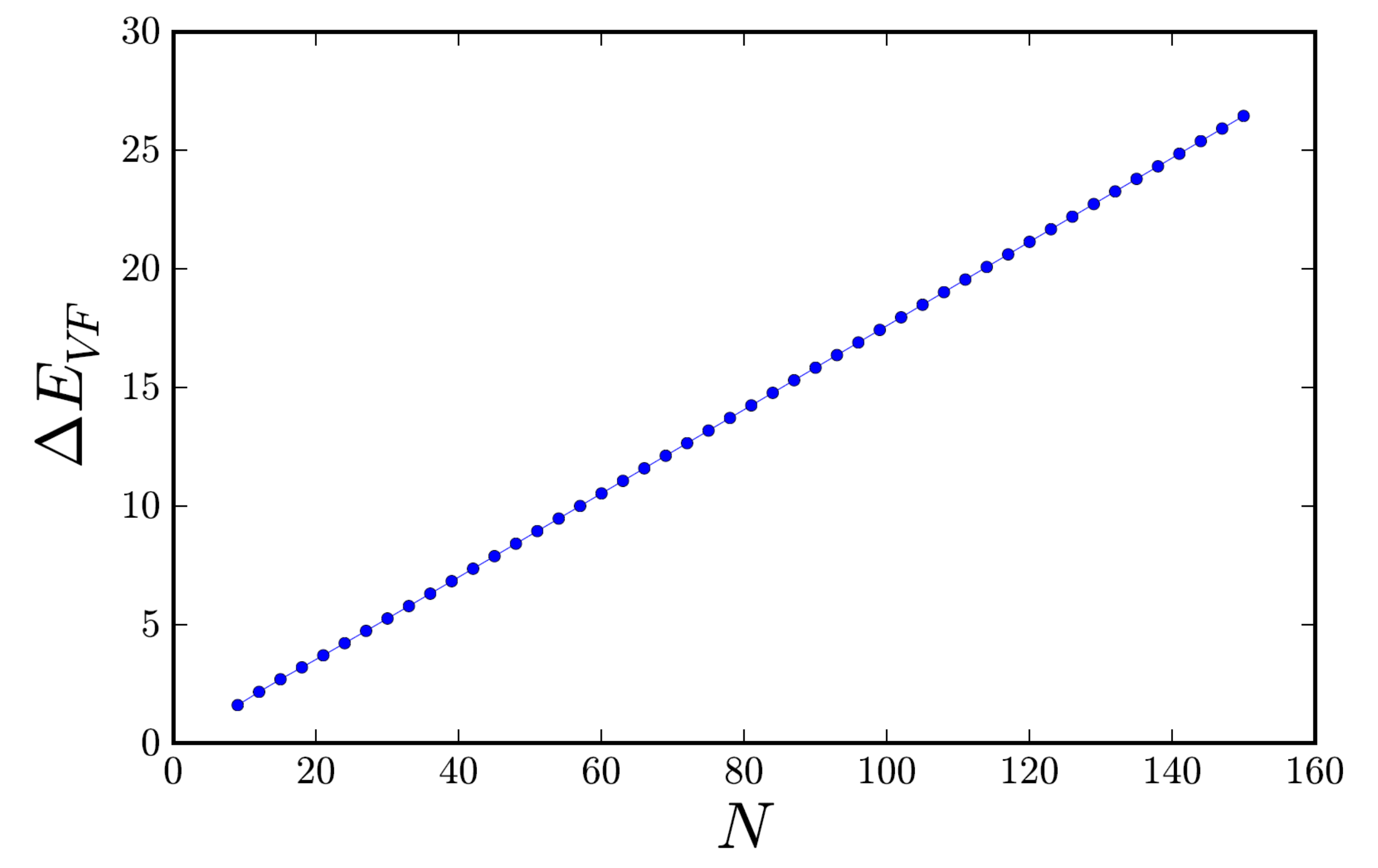}
	\caption{Energy $E_{VF}$ of the vortex-full sector as a function of $N$ with $J_{x}=1.0$, $J_{y}=-0.55$, $J_{z}=0.25$ in $\mathcal{S}_{1,3}$, while $J_{z'}=4$ in $\mathcal{S}_{2}$. As expected, the energy of the vortex-full sector is always positive and grows linearly with $N$. The slope of the of the straight line can be interpreted as an average vortex energy. This plot indicates again that the vortex-vortex interaction does not favor the creation of vortices and the ground state is vortex-free. The junction between sections $\mathcal{S}_{1}$ and $\mathcal{S}_{2}$ is at $p=2N/3$ and between $\mathcal{S}_{2}$ and $\mathcal{S}_{3}$ is at $p=4N/3$.}
	\label{fig:Vortex_Energy_supp_5}
\end{figure}

As discussed in the main text, although Lieb's theorem \cite{Lieb1994,Macris1996} is not directly applicable to our system, we nevertheless are able to show numerically that the  ground state is indeed vortex-free for $\m{sgn}(J_{x})=-\m{sgn}(J_{y})$ and $J_{z_{ij}}>0$, while it is vortex-full for $\m{sgn}(J_{x})=\m{sgn}(J_{y})$. Let us focus on the case  $\m{sgn}(J_{x})=-\m{sgn}(J_{y})$, since the other one can easily be deduced from it as discussed in the main text. Figure~\ref{fig:Vortex_energy_supplement_1} shows single-vortex energies for different ladder lengths and $J_{x,y,z,z'}$ coupling configurations. All the results are consistent with our assumption that the ground state is vortex-free. We have furthermore investigated the effect of vortex-vortex interactions and plotted in Figs.~\ref{fig:Vortex_energy_supplement_4_1} and \ref{fig:Vortex_energy_supplement_4_2} multi vortex energies for different $N$, $J_{x,y,z}$ couplings, and distance between the vortices. Again, all the plots indicate a vortex-free ground state since the \textit{attractive} vortex-vortex interaction is not strong enough to favor the creation of additional vortices. 
Finally, we plot in Fig.~\ref{fig:Vortex_Energy_supp_5} the energy of the vortex-full sector as a function of  $N$ for $J_{x}=1.0$, $J_{y}=-0.55$, $J_{z}=0.25$ in $\mathcal{S}_{1,3}$, while $J_{z'}=4$ in $\mathcal{S}_{2}$. The energy of the vortex-full sector is always positive and increases linearly with the system size $N$. This result again shows that vortex-vortex interactions do not favor the creation of vortices and the ground state is free of vortices. We have checked that this result is valid for many other choices of parameters $J_{x,y,z}$. A detailed explanation of the plots is given in the figure captions.

As a final remark, we would like to mention that the groundstate of the trijunction setup presented in the main text (see Fig.~5 of the main text) is also vortex-free or -full. Indeed, each ladder forming the trijunction is separately free or full of vortices and thus by a continuity argument it is clear that switching on (small) couplings between different ladders cannot create vortices.

\section{Different mapping to study the robustness of MES}\label{sec:newmapping}
The aim of this Appendix is to propose another mapping to study the robustness of MES in the homogeneous and inhomogeneous ladder. We use the same labeling as Fig.~\ref{fig:Majorana_S_W} but now make use of the following spin-to-spin mapping: \cite{Fradkin1978,Feng2007}
\begin{eqnarray}
\sigma_{i}^{z}&\rightarrow& \tau_{i-1}^{z}\tau_{i}^{z},\\
\sigma_{i}^{x}&\rightarrow& \prod_{k=i}^{4N}\tau_{k}^{x},
\end{eqnarray}
where $\tau_{i}^{x,y,z}$ are usual Pauli matrices. The mapping is thus to an auxiliary system of $4N+1$ $\tau_{i}$ spins ($i=0,1,2,...,4N$). Therefore, the mapping introduces a doubling of the Hilbert space which can be taken into account by imposing the constraint $\tau_{0}^{z}=+1$.

The spin ladder Hamiltonian (\ref{eq:Hamiltonian}) takes in this new language the following form:
\begin{eqnarray}
H&=&J_{x}\left(\tau_{2}^{x}+\tau_{4}^{x}+...+\tau_{N-2}^x\right)\nonumber\\
&&+J_{z}(\underbrace{\tau_{0}^{z}}_{+1}\tau_{2}^{z}+\tau_{2}^{z}\tau_{4}^z+\tau_{4}^z\tau_6^z+...+\tau_{N-2}^z\tau_{N}^z)\nonumber\\
&&+J_y(\tau_{0}^z\tau_1^y\tau_2^x\tau_3^y\tau_4^z+...+\tau_{N-4}^z\tau_{N-3}^y\tau_{N-2}^x\tau_{N-1}^y\tau_{N}^z).\nonumber\\
\end{eqnarray}
This Hamiltonian possesses many conserved quantities since all $\tau_{i}^{y}$ with odd $i$ commute with $H$ and are straightforwardly connected to the vortex operators:
\begin{equation}
W_n=-\tau_{2n-1}^y\tau_{2n+1}^y\,\,\,\,\,\,\m{and}\,\,\,\,\,\,\bar{W}_n=-\tau_{2n+1}^y\tau_{2n+3}^y. 
\end{equation}
\subsection{Homogeneous ladder}
As a first step, let us now study the homogeneous ladder and consider the case where $J_{x}\gg J_{y,z}$ which, according to Eq.~(\ref{eq:nu}), is in the topological phase. For $J_{y,z}=0$, the odd spins are completely decoupled and the corresponding $2^{N/2-1}$ degeneracy is due to the absence of gap to create a vortex. However, there is an additional and interesting degeneracy coming from the fact that the last spin $\tau_N^x$ does not appear in the Hamiltonian and the total degeneracy in the topological phase is thus $2^{N/2}$. In the opposite limit, i.e., when $J_z\neq 0$ and $J_{x,y}=0$, which, according to Eq.~(\ref{eq:nu}), corresponds to the nontopological phase, then by construction $\tau_{0}^z=+1$ and no additional degeneracy is present. In this limit the degeneracy is thus $2^{N/2-1}$. We thus identify the additional degeneracy due to the $N^{\m{th}}$ spin as the topological degeneracy. In the strong $J_x$ limit, the two corresponding degenerate ground states take the following form:
\begin{equation}
\vert\psi_{\uparrow\downarrow}\rangle=\vert \uparrow_{0} +_2 +_4....+_{N-2} \uparrow\downarrow_N\rangle,
\end{equation}
where $+_i$ is an eigenstate of $\tau_i^x$ while $\uparrow_j$ and $\downarrow_j$ are eigenstates of $\tau_j^z$.

It is now straightforward to understand that the single-body perturbation $V=\sigma_{N}^{x}=\tau_{N}^x$ will split the topological degeneracy since
\begin{equation}
\langle\psi_\uparrow\vert V\vert\psi_\downarrow\rangle=\langle\psi_\uparrow\vert \tau_N^x\vert\psi_\downarrow\rangle\neq0.
\end{equation}
With the Kitaev's mapping, we noticed the presence of an additional degeneracy because of the presence of six MES in the extended space. In the new language, this additional degeneracy corresponds to the sign of $\tau_{1}^{y}$ (the state with all $\tau_{2i-1}^{y}=+1$ is degenerate with the state where all $\tau_{2i-1}^{y}=-1$). This degeneracy is split by $\sigma_{1}^{x}=\prod_{i=1}^{N}\tau_{i}^{x}$ since it induces transition between the states with all $\tau_{2i-1}^{y}=+1$ and all $\tau_{2i-1}^{y}=-1$.
\subsection{Inhomogeneous ladder}
\begin{figure}[h]
	\centering
		\includegraphics[width=0.48\textwidth]{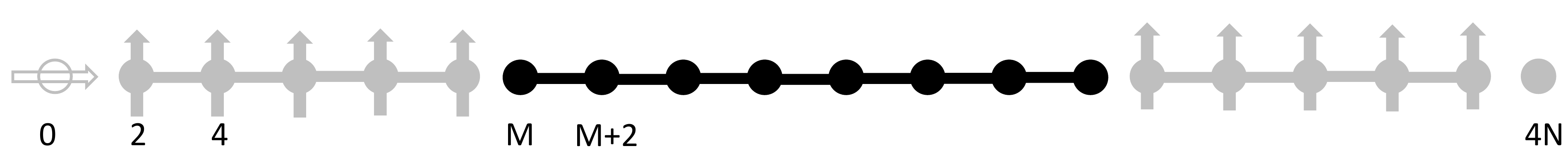}
	\caption{Pictorial representation of the groundstates in the limit $J_{x}\neq0$ and $J_{y,z}=0$ in sections $\mathcal{S}_{1,3}$ and $J_{z}\neq0$ and $J_{x,y}=0$ in $\mathcal{S}_{2}$. Here $\Uparrow$ and $\Downarrow$ are a pictorial representations of $\tau^x$ eigenstates.}
	\label{fig:GroundState}
\end{figure}

Let us now consider the inhomogeneous ladder with two topological sections $\mathcal{S}_{1,3}$ separated by the nontopological section $\mathcal{S}_2$, see Fig.~\ref{fig:Chain_5}. In the case where $J_x\neq0$ and $J_{y,z}=0$ in $S_{1,3}$, while $J_z\neq0$ and $J_{x,y}=0$ in $S_2$ it is easy to see the origin of the eightfold degeneracy, see Fig.~\ref{fig:GroundState}. One degeneracy is related to the sign of $\tau_{1}^y=\pm 1$. This degeneracy is easy to split with a local perturbation $\epsilon\tau_{1}^{y}=\epsilon\sigma_{1}^y\sigma_{2}^x$ [as we showed with Kitaev's mapping (\ref{eq:mapping}), $\sigma_{1}^{x}=\prod_{i=1}^{N}\tau_{i}^{x}$ also splits this degeneracy].
The origin of the remaining fourfold degeneracy can be understood easily in the limiting case we consider. The first twofold degeneracy is due to the $N^{\m{th}}$ spin which is decoupled from the other spins while the other twofold degeneracy comes from the middle Ising section, which has the same energy if all spins are up or down. The former degeneracy is trivially split by a perturbation $\epsilon\tau_{N}^x=\epsilon\sigma_{N}^{x}$, while the later is clearly split by the perturbation in Eq.~(\ref{eq:twobodyper}),
\begin{eqnarray}
\epsilon\sigma_{M-1}^{x}\sigma_{M}^y&=&\epsilon i\sigma_{M-1}^x \sigma_{M}^x\sigma_{M}^z\\
&=&\epsilon i \tau_{M-1}^x\tau_{M-1}^z\tau_{M}^z=\epsilon\tau_{M-1}^y
\tau_{M}^z,
\end{eqnarray}
since $\tau_{M-1}^y$ is a conserved quantity ($M-1$ is an odd site here).
We have thus shown here explicitly that two-body perturbations are enough to split the remaining degeneracies in agreement with the discussion in the main text. 
\color{black}

\section{Proliferation of $\pi$-junction zero-modes in $XX$-$YY$ spin chain}\label{sec:proliferation}
In this section we study some properties of the zero-energy modes present in a $xx$-$yy$ spin chain of length $2N$, described by the Hamiltonian
\begin{equation}
H_{xx-yy}=J_{x}\sum_{i\,\m{odd}}^{2N-1}\sigma_{i}^{x}\sigma_{i+1}^{x}+J_{y}\sum_{i\,\m{even}}^{2N}\sigma_{i}^{y}\sigma_{i+1}^{y}.
\end{equation}
We show that the $xx$-$yy$ spin chains contain many additional zero-energy modes besides the two $c$ Majoranas localized at the ends of the chain. In the language of mapping (\ref{eq:mapping}), this arises because all non equivalent $u$ configurations are degenerate; that is, putting a $u_{ij}^{\alpha}=-1$ does not cost energy. It is instructive to study this model with a Jordan-Wigner transformation
\begin{equation}\label{eq:Jordan}
\sigma_{j}^{+}=\prod_{k=1}^{j-1}(-1)^{n_{k}}a_{j}\,\,\,\,\m{and}\,\,\,\,\sigma_{j}^{-}=\prod_{k=1}^{j-1}(-1)^{n_{k}}a_{j}^{\dagger},
\end{equation}
where $a_{j}$ annihilates a complex fermion at site $j$, i.e., $\{a_{j}^{(\dagger)},a_{j'}^{(\dagger)}\}=0$ and $\{a_{j},a_{j'}^{\dagger}\}=\delta_{jj'}$, and $n_{j}=a_{j}^{\dagger}a_{j}$.
\color{black}
With the use of Eq. (\ref{eq:Jordan}),  $H_{xx-yy}$ takes the  form
\begin{eqnarray}\label{eq:HamiltonianJordan}
\widetilde{H}_{xx-yy}&=&\sum_{i\,\m{odd}}\left(-w_{x}a_{i}^{\dagger}a_{i+1}+\Delta_{x}a_{i}a_{i+1}+\m{h.c.}\right)\nonumber\\
&+&\sum_{i\,\m{even}}\left(-w_{y}a_{i}^{+}a_{i+1}+\Delta_{y}a_{i}a_{i+1}+\m{h.c.}\right),\nonumber\\
\end{eqnarray}
where $w_{x}=\Delta_{x}=J_{x}/4$ and $w_{y}=-\Delta_{y}=-J_{y}/4$.

\begin{figure}[h]
	\centering
		\includegraphics[width=0.40\textwidth]{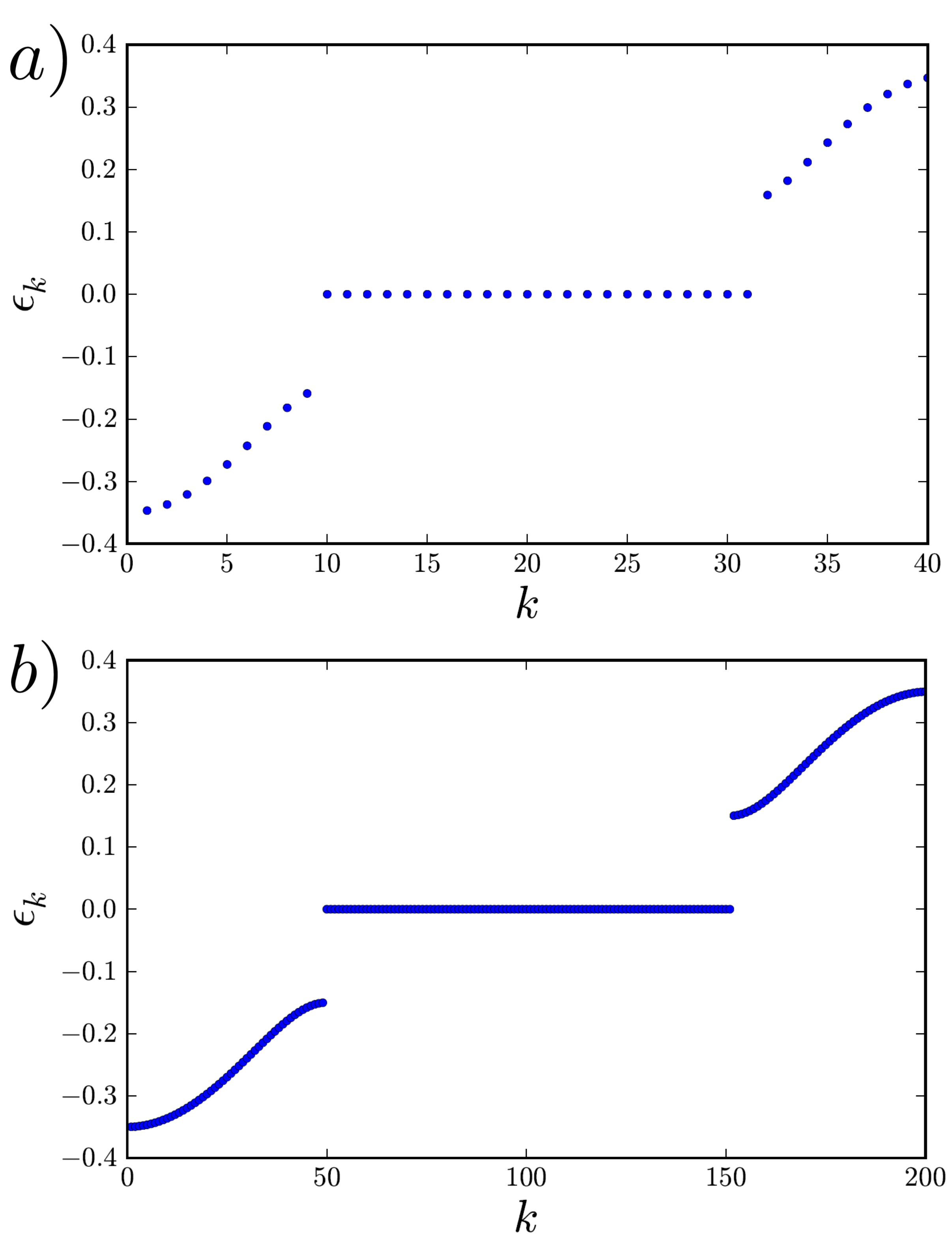}
	\caption{Energy eigenvalues $\epsilon_{k}$ of $\mathcal{H}$ in Eq.~(\ref{eq:kl}) for $xx$-$yy$ chains with $J_{x}=0.4$, $J_{y}=1.0$, and $2N=20$ for $a)$ and $2N=100$ for $b)$. There are $2N$ zero-energy modes in addition to the two expected MES. The presence of these additional zero modes can be understood by mapping Hamiltonian $H_{xx-yy}$ to an array of $\pi$ junctions [see Eq.~\ref{eq:HamiltonianJordan}].}
	\label{fig:xx_yy_chain_spectrum_1}
\end{figure}

Since there is a difference of $\pi$ in the pase of $\Delta_{x}$ and $\Delta_{y}$, we thus conclude that Hamiltonian (\ref{eq:HamiltonianJordan}) represents an array of $\pi$ junctions and thus possesses $2N$ additional zero-energy modes. \cite{AliceaWire2011} To find the spectrum $\epsilon_{k}$ of Hamiltonian (\ref{eq:HamiltonianJordan}), we artificially double the number of degrees of freedom and rewrite Eq.~(\ref{eq:HamiltonianJordan}) as
\begin{equation}\label{eq:kl}
\widetilde{H}_{xx-yy}=\frac{1}{2}{\bf a}\,\mathcal{H}\,{\bf a}^{\dagger},
\end{equation}
where  ${\bf a}=\begin{pmatrix}a_{1} & ... & a_{2N} & a_{1}^{\dagger} & ... & a_{2N}^{\dagger}\end{pmatrix}$, and $\mathcal{H}$ is a real $4N\times 4N$ symmetric matrix defined through Eq.~(\ref{eq:HamiltonianJordan}).
Figure~\ref{fig:xx_yy_chain_spectrum_1} is a plot of the eigenvalues $\epsilon_{k}$ of $\mathcal{H}$ which corresponds to the modes of a $xx$-$yy$ spin chain with $J_{x}=0.4$, $J_{y}=1.0$, and length $2N=20$ for $a)$ and $2N=100$ for $b)$. As expected, the number of additional zero-energy modes is $2N$. It is thus possible to generate zero-energy modes in the $xx$-$yy$ spin chain by increasing the system size.
\section{Long-distance spin-spin correlation function}\label{sec:correlations}

\begin{figure}[h!]
	\centering
		\includegraphics[width=0.36\textwidth]{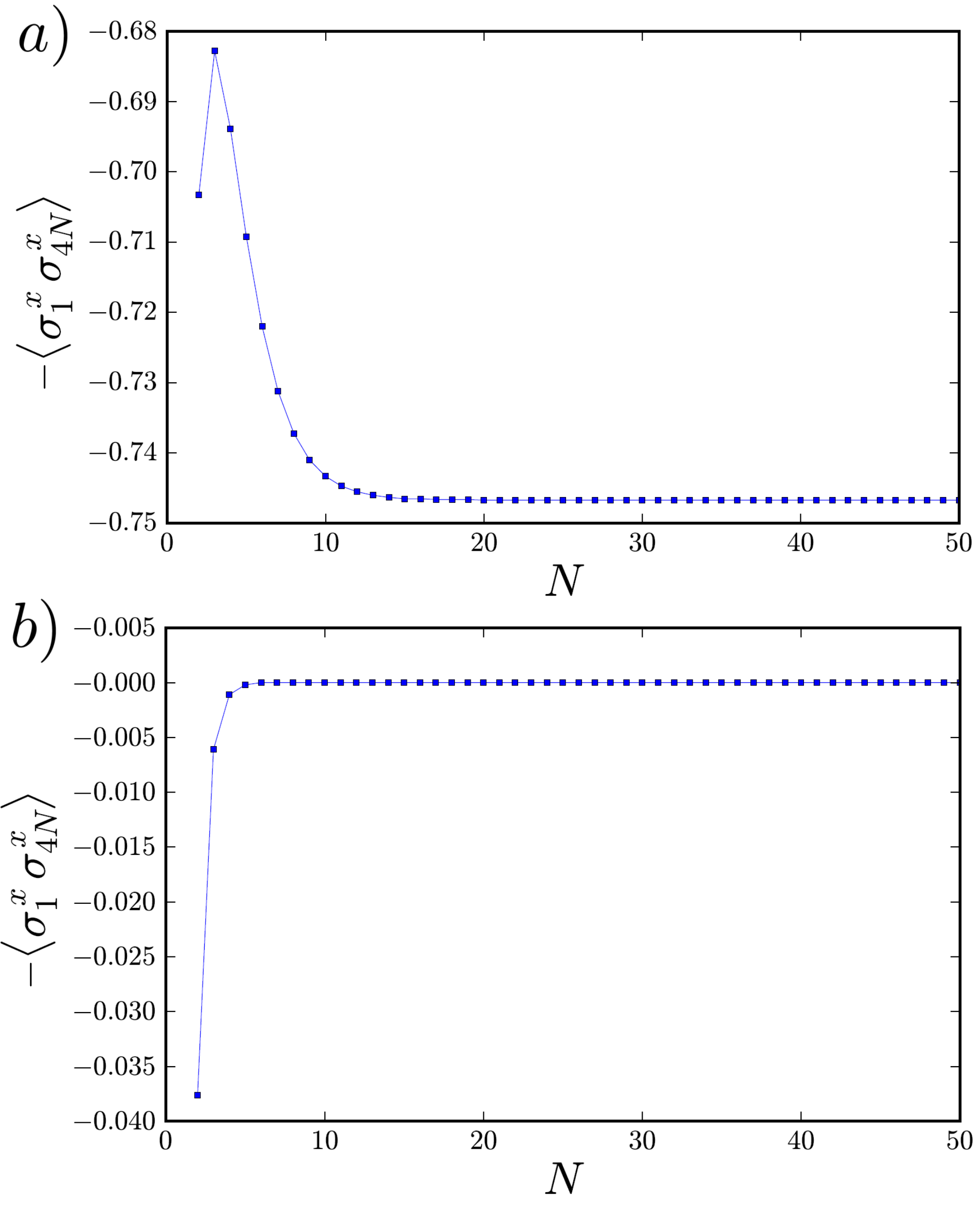}
	\caption{$-\langle\Psi_{n}\vert\sigma_{1}^{x}\sigma_{4N}^{x}\vert\Psi_{n}\rangle$ as a function of $N$, with all $u_{ij}=+1$, $J_{x}=1.0$, $J_{y}=-0.4$, and $J_{z_{1}}=J_{z_{2}}=J_{z_{3}}=0.2$ for $a)$, and $J_{z_{1}}=J_{z_{2}}=J_{z_{3}}=2$ for $b)$. We make use of the projection protocol of Ref.~\onlinecite{Chesi2011} in order to determine if the physical ground state of the vortex-free sector has even ($n=0$) or odd ($n=1$) parity.}
	\label{fig:Correlations_N_jx=10_jy=-04_jz=02}
\end{figure}

In this section we study the static long-distance spin-spin correlation function
$\langle\sigma_{1}^{x}\sigma_{4N}^{x}\rangle$ (the site labeling is shown in Fig.~3 of the main text). We note that this correlator vanishes in the standard honeycomb model \cite{Shankar2007,Nussinov2008} but is non zero for the ladder in the topological phase due to the presence of MES localized at sites $1$ and $4N$ when $J_{x}>J_{y}$ (the scenario with $J_{x}<J_{y}$ can be treated analogously by considering $\langle\sigma_{2}^{y}\sigma_{4N-1}^{y}\rangle$). Let us first give an explicit expression for $\langle\sigma_{1}^{x}\sigma_{4N}^{x}\rangle$.
\begin{widetext}
Since
\begin{equation}
\sigma_{i}^{x}\sigma_{j}^{x}=-i u_{ij}^{x}c_{i}c_{j}
\end{equation}
and
\begin{equation}
(b_{1}, b_{2},...,b_{4N-1},b_{4N})Q^{u}=(c_{1},...,c_{4N}),
\end{equation}
we  have
\begin{equation}
c_{i}=\sum_{k}^{N}Q_{ki}^{u}b_{k}
\end{equation}
\begin{eqnarray}
c_{i}c_{j}=\sum_{k,k'}Q_{ki}^{u}Q_{k'j}^{u}b_{k}b_{k'}.
\end{eqnarray}
Using
\begin{eqnarray}
a_{k}^{\dagger}&=&(b_{2k-1}-ib_{2k})/2\nonumber\\
a_{k}&=&(b_{2k-1}+ib_{2k})/2,
\end{eqnarray}
and
\begin{equation}
c_{i}c_{j}=\sum_{l,k}Q_{2k-1i}^{u}Q_{2l-1j}^{u}b_{2k-1}b_{2l-1}+\sum_{l,k}Q_{2k-1i}^{u}Q_{2lj}^{u}b_{2k-1}b_{2l}+\sum_{l,k}Q_{2ki}^{u}Q_{2l-1j}^{u}b_{2k}b_{2l-1}^{'}+\sum_{l,k}Q_{2ki}^{u}Q_{2lj}^{u}b_{2k}b_{2l}\, ,
\end{equation}
we obtain
\begin{eqnarray}
c_{i}c_{j}&=&\sum_{l,k}Q_{2k-1i}^{u}Q_{2l-1j}^{u}(a_{k}+a_{k}^{\dagger})(a_{l}+a_{l}^{\dagger})+\sum_{l,k}Q_{2k-1i}^{u}Q_{2lj}^{u}(a_{k}+a_{k}^{\dagger})(1/i)(a_{l}-a_{l}^{\dagger})\nonumber\\
&&+\sum_{l,k}Q_{2ki}^{u}Q_{2l-1j}^{u}(1/i)(a_{k}-a_{k}^{\dagger})(a_{l}+a_{l}^{\dagger})+\sum_{l,k}Q_{2ki}^{u}Q_{2lj}^{u}(1/i^2)(a_{k}-a_{k}^{\dagger})(a_{l}-a_{l}^{\dagger})\, ,
\end{eqnarray}
and thus
\begin{eqnarray}
c_{i}c_{j}&=&\sum_{k,l}\left[Q_{2k-1i}^{u}Q_{2l-1j}^{u}(a_{k}a_{l}+a_{k}a_{l}^{\dagger}+a_{k}^{\dagger}a_{l}+a_{k}^{\dagger}a_{l}^{\dagger})\right.\nonumber\\
&&\left.+Q_{2k-1i}^{u}Q_{2lj}^{u}(1/i)(a_{k}a_{l}-a_{k}a_{l}^{\dagger}+a_{k}^{\dagger}a_{l}-a_{k}^{\dagger}a_{l}^{\dagger})\right.\nonumber\\
&&\left.+Q_{2ki}^{u}Q_{2l-1j}^{u}(1/i)(a_{k}a_{l}+a_{k}a_{l}^{\dagger}-a_{k}^{\dagger}a_{l}-a_{k}^{\dagger}a_{l}^{\dagger})\right.\nonumber\\
&&\left.+Q_{2ki}^{u}Q_{2lj}^{u}(-a_{k}a_{l}+a_{k}a_{l}^{\dagger}+a_{k}^{\dagger}a_{l}-a_{k}^{\dagger}a_{2l}^{\dagger})\right] \, .
\end{eqnarray}
It is now straightforward to calculate $\langle\Psi_{n=0}\vert\sigma_{1}^{x}\sigma_{4N}^{x}\vert\Psi_{n=0}\rangle$, where $n$ ($n=0,1$) represents the filling of MES while all the high-energy modes are unfilled,
\begin{eqnarray}
\langle\Psi_{n=0}\vert\sigma_{1}^{x}\sigma_{4N}^{x}\vert\Psi_{n=0}\rangle&=&-i u_{14N}^{x}\langle n=0\vert c_{1}c_{4N}\vert n=0\rangle\nonumber\\
&=&-i u_{14N}^{x}\langle n=0\vert \sum_{k}\left[Q_{2k-1i}^{u}Q_{2k-1j}^{u}(a_{2k-1}a_{2k-1}^{\dagger}+a_{2k-1}^{\dagger}a_{2k-1})+Q_{2k-1i}^{u}Q_{2kj}^{u}(1/i)(-a_{k}a_{k}^{\dagger}+a_{k}^{\dagger}a_{k})\right.\nonumber\\
&&\left.+Q_{2ki}^{u}Q_{2k-1j}^{u}(1/i)(a_{k}a_{k}^{\dagger}-a_{k}^{\dagger}a_{k})+Q_{2ki}^{u}Q_{2kj}^{u}(a_{k}a_{k}^{\dagger}+a_{k}^{\dagger}a_{k})\right]\vert n=0\rangle,
\end{eqnarray}
where $i=1$ and $j=4N$.

With the use of the fermionic anticommutation relation $\{a_{k},a_{k}^{\dagger}\}=1$ we obtain
\begin{eqnarray}
\langle\Psi_{n=0}\vert\sigma_{1}^{x}\sigma_{4N}^{x}\vert\Psi_{n=0}\rangle&=&-i u_{14N}^{x}\sum_{k}\left[Q_{2k-1i}^{u}Q_{2k-1j}^{u}-(1/i) Q_{2k-1i}^{u}Q_{2kj}^{u}+(1/i)Q_{2ki}^{u}Q_{2k-1j}^{u}+Q_{2ki}^{u}Q_{2kj}^{u}\right]\nonumber\\
&&+\sum_{k}\left[(2/i) Q_{2k-1i}^{u}Q_{2kj}^{u}-(2/i) Q_{2ki}^{u}Q_{2k-1j}^{u}\right](n_{k}=0)\\
&=&-i u_{14N}^{x}\sum_{k}\left[Q_{2k-1i}^{u}Q_{2k-1j}^{u}-(1/i) Q_{2k-1i}^{u}Q_{2kj}^{u}+(1/i)Q_{2ki}^{u}Q_{2k-1j}^{u}+Q_{2ki}^{u}Q_{2kj}^{u}\right]\, .
\end{eqnarray}
Since the matrix $Q^{u}$ is orthogonal we finally obtain
\begin{eqnarray}
\langle\Psi_{n=0}\vert\sigma_{1}^{x}\sigma_{4N}^{x}\vert\Psi_{n=0}\rangle&=&-i u_{14N}^{x}\sum_{k}\left[-(1/i) Q_{2k-1i}^{u}Q_{2kj}^{u}+(1/i)Q_{2ki}^{u}Q_{2k-1j}^{u}\right]\\
&=&u_{14N}^{x}\sum_{k}\left[ Q_{2k-1i}^{u}Q_{2kj}^{u}-Q_{2ki}^{u}Q_{2k-1j}^{u}\right]\label{eq:minuscorr1},
\end{eqnarray}
where we recall that $u_{14N}^{x}=\pm1$ decouples from the Hamiltonian in the absence of external perturbations.

Similarly we can show that
\begin{equation}\label{eq:minuscorr2}
\langle\Psi_{n=1}\vert\sigma_{1}^{x}\sigma_{4N}^{x}\vert\Psi_{n=1}\rangle=u_{14N}^{x}(\sum_{k}\left[ Q_{2k-1i}^{u}Q_{2kj}^{u}-Q_{2ki}^{u}Q_{2k-1j}^{u}\right]+2Q_{2\alpha i}^{u}Q_{2\alpha-1j}^{u}-2Q_{2\alpha-1i}^{u}Q_{2\alpha j}^{u}),
\end{equation}
\end{widetext}
where $\alpha$ is the index of the fermonic mode formed by the Majoranas; that is, $n_{\alpha}=n=1$ is the filling of MES.
As mentioned above, the long-distance spin-spin correlation $\langle\sigma_{1}^{x}\sigma_{4N}^{x}\rangle$ vanishes in the standard honeycomb model \cite{Shankar2007,Nussinov2008} and is non zero here only in the topological phase due to the presence of the MES state with components on both sites $1$ and $4N$. We show in Figs.~\ref{fig:Correlations_N_jx=10_jy=-04_jz=02} $a)$ and $b)$ a plot of $-\langle\Psi_{n}\vert \sigma_{1}^{x}\sigma_{4N}^{x}\vert\Psi_{n}\rangle$ as a function of $N$ with all $u_{ij}=+1$, $J_{x}=1.0$, $J_{y}=-0.4$, and $J_{z_{1}}=J_{z_{2}}=J_{z_{3}}=0.2$ for the topological phase in $a)$ and $J_{z_{1}}=J_{z_{2}}=J_{z_{3}}=2$ for the nontopological phase in $b)$. We make use of the projection protocol of Ref.~\onlinecite{Chesi2011} in order to determine if the physical ground state of the vortex-free sector has even ($n=0$) or odd ($n=1$) parity. As expected, the long-distance spin-spin correlation takes a finite value in $a)$ while it vanishes in $b)$. 

\begin{figure}[h!]
	\centering
		\includegraphics[width=0.36\textwidth]{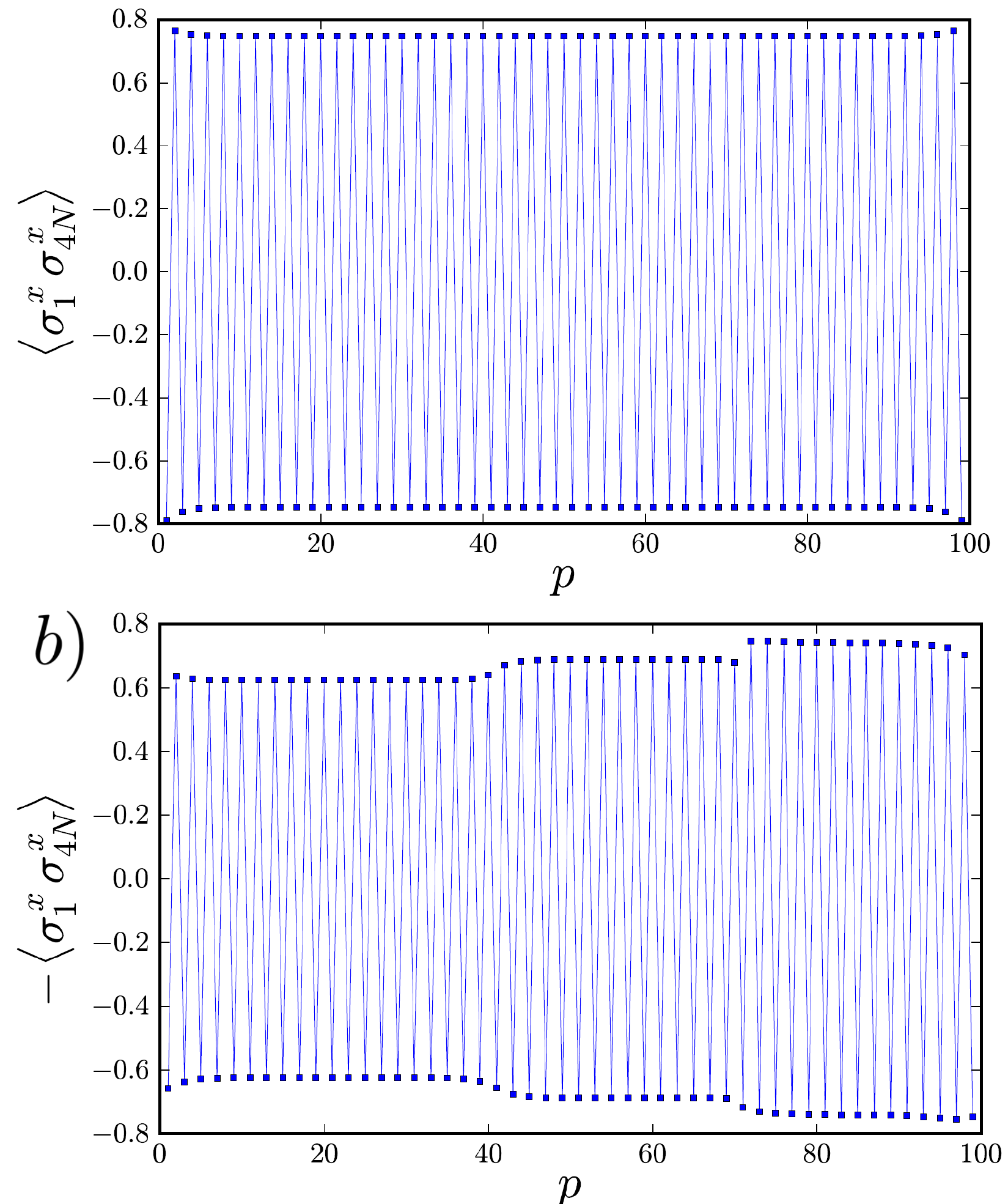}
	\caption{Plot of correlator $-\langle\Psi_{n}\vert\sigma_{1}^{x}\sigma_{4N}^{x}\vert\Psi_{n}\rangle$ as a function of  position of a single vortex, $p$, for $N=50$, $J_{x}=1.0$, $J_{y}=-0.4$, and $J_{z_{1}}=J_{z_{2}}=J_{z_{3}}=0.2$ in $a)$ and $J_{z_{1}}=J_{z_{3}}=0.2$, $J_{z_{2}}=2$ in $b)$. The junctions between sections $\mathcal{S}_{1,3}$ and $\mathcal{S}_{2}$ are at plaquettes $p=41,71$. We used the projection protocol of Ref.~\onlinecite{Chesi2011} to determine if the physical ground state of the corresponding single-vortex sector has even ($n=0$) or odd ($n=1$) parity. Note that the physical groundstates we consider have a fixed parity $i\gamma_{2}\gamma_{3}=+1$ and oscillating parity $i\gamma_{1}\gamma_{4}$.}
	\label{fig:Correlations_p_jx=10,jy=-04,jz=02}
\end{figure}

In the remainder of this section, we want to investigate the effects of vortices on the long-distance correlation function. Figure~\ref{fig:Correlations_p_jx=10,jy=-04,jz=02}$a)$ shows a plot of $-\langle\Psi_{n}\vert\sigma_{1}^{x}\sigma_{4N}^{x}\vert\Psi_{n}\rangle$ as a function of  position of a single vortex, $p$, for $N=50$, $J_{x}=1.0$, $J_{y}=-0.4$, $J_{z_{1}}=J_{z_{2}}=J_{z_{3}}=0.2$. This ladder has one topological section with two MES $\gamma_{1,2}$ localized on the left and right ends. The oscillations between positive and negative values of the correlator show that the vortex changes the value of $\langle\Psi_{n}\vert\sigma_{1}^{x}\sigma_{4N}^{x}\vert\Psi_{n}\rangle$ and thus the MES parity $i\gamma_{1}\gamma_{2}$ as a function of its position on the ladder. Indeed, using Eqs.~(\ref{eq:minuscorr1}) and (\ref{eq:minuscorr2}) we show numerically that 
$\langle \Psi_{n=0}\vert\sigma_{1}^{x}\sigma_{4N}^{x}\vert\Psi_{n=0}\rangle =-\langle\Psi_{n=1}\vert\sigma_{1}^{x}\sigma_{4N}^{x}\vert\Psi_{n=1}\rangle$, and thus conclude that
a change of sign in the correlator implies a change of the parity $i\gamma_{1}\gamma_{2}$ ({\it i.e.},  $n=0\leftrightarrow n=1$).
We make use of the projection protocol of Ref.~\onlinecite{Chesi2011} in order to determine if the physical ground state of the one-vortex sectors have even ($n=0$) or odd ($n=1$) parity. In Fig.~\ref{fig:Correlations_p_jx=10,jy=-04,jz=02} $b)$ we plot  $-\langle\Psi_{n}\vert\sigma_{1}^{x}\sigma_{4N}^{x}\vert\Psi_{n}\rangle$ as a function of position of a single vortex, $p$, for $N=50$, $J_{x}=1.0$, $J_{y}=-0.4$, $J_{z_{1}}=J_{z_{3}}=0.2$, $J_{z_{2}}=2$. This ladder carries four MES: $\gamma_{1,4}$ at respectively the left and right ends of the ladder and $\gamma_{2,3}$ at the junction between topological and nontopological sections. The oscillations in the correlator demonstrate again oscillations in the parity $i\gamma_{1}\gamma_{4}$. As mentioned in the main text, the one-vortex state is highly degenerate since it does not cost energy to move a vortex to a nearby plaquette, and thus, without any prior measurement, the position of a vortex is generally not known, and so neither is the parity of the MES. In the case where MES are used for topological computing (braiding), it is thus essential that the ground state is vortex-free or vortex-full.
\color{black}

\end{document}